# von Willebrand Factor unfolding mediates platelet deposition in a model of high-shear thrombosis

Mansur Zhussupbekov[1], Rodrigo Méndez Rojano[1], Wei-Tao Wu[2], James F. Antaki[1,*]

[1] Meinig School of Biomedical Engineering, Cornell University, Ithaca, NY, USA
[2] Department of Aerospace Science and Technology, Nanjing University of Science and Technology, Nanjing, China

*Correspondence: antaki@cornell.edu

## ABSTRACT
Thrombosis under high-shear conditions is mediated by the mechanosensitive blood glycoprotein von Willebrand Factor (vWF). vWF unfolds in response to strong flow gradients and facilitates rapid recruitment of platelets in flowing blood. While the thrombogenic effect of vWF is well recognized, its conformational response in complex flows has largely been omitted from numerical models of thrombosis. We recently presented a continuum model for the unfolding of vWF, where we represented vWF transport and its flow-induced conformational change using convection-diffusion-reaction equations. Here, we incorporate the vWF component into our multi-constituent model of thrombosis, where the local concentration of stretched vWF amplifies the deposition rate of free-flowing platelets and reduces the shear cleaning of deposited platelets. We validate the model using three benchmarks: in vitro model of atherothrombosis, a stagnation point flow, and the PFA-100®, a clinical blood test commonly used for screening for von Willebrand Disease (vWD). The simulations reproduced the key aspects of vWF-mediated thrombosis observed in these experiments, such as the thrombus location, thrombus growth dynamics, and the effect of blocking platelet-vWF interactions. The PFA-100® simulations closely matched the reported occlusion times for normal blood and several hemostatic deficiencies, namely, thrombocytopenia, vWD Type 1, and vWD Type 3. Overall, this multi-constituent model of thrombosis enables macro-scale 3-D simulations of thrombus formation in complex geometries over a wide range of shear rates and accounts for qualitative and quantitative hemostatic deficiencies in patient blood.

## SIGNIFICANCE
Application of numerical models of thrombosis to the problem of arterial clot formation revealed that to achieve platelet aggregation at pathologically high shear rates, platelet adhesion rates or adhesion strength must be increased to account for platelet-vWF interactions. Many existing models have correlated platelet adhesion parameters to local shear rate or elongation rate, but not both. vWF, on the other hand, exhibits different responses in simple shear and extensional flow, with distinct thresholds for unfolding and conformational hysteresis effects. The multi-constituent model of thrombosis presented in this work features our continuum model of vWF unfolding and demonstrates its utility in predicting macro-scale thrombus formation in high-shear conditions. The vWF component could also be adapted to other numerical models of thrombosis.

## INTRODUCTION
von Willebrand Factor (vWF), a large multimeric blood glycoprotein, is essential for normal hemostasis and plays a critical role in life-threatening scenarios such as arterial bleeding (1, 2). Upon blood vessel injury, subendothelial vWF is exposed and plasma vWF binds to collagen – all acting as an adhesive substrate for flowing platelets. Rapid electrostatically steered binding between vWF A1 domain and platelet membrane receptor GPIbα enables platelet recruitment even in fast-flowing blood (3), and the increased number of available binding sites on vWF concatemers facilitates the capture of millions of platelets within minutes (4, 5).

This innate hemostatic capacity of the vWF is regulated by its conformational response to flow: vWF normally circulates in plasma in a compact globular configuration, but will unfold into a long flexible chain when subjected to strong shear or extensional flow gradients that occur in a bleeding vessel (6–8). Tensile force acting on an extended chain exposes and activates the vWF-A1 domain that binds to collagen and GPIbα on the platelet membrane, but also exposes the A2 domain that is required for enzymatic cleavage and size regulation of vWF concatemers (9–12). While these force-dependent regulation mechanisms normally restrict the action of vWF to the site of injury, pathological conditions such as atherosclerotic plaque rupture can unleash the thrombogenic potential of vWF and lead to occlusive thrombus formation (13–15).

vWF-mediated platelet accumulation at sites of atherosclerotic plaque rupture is one of the key events in the pathophysiology of acute ischemic stroke and myocardial infarction (16, 17). Histological analyses of thrombi retrieved from ischemic stroke patients detected abundant presence of vWF collocated with dense fibrin structures and platelet aggregates (18, 19). Numerous in vivo and in vitro experiments featuring stenotic geometries that mimic atherosclerotic plaques have demonstrated the key role of vWF in mediating high-shear thrombus formation (14, 20–24).

Accordingly, application of numerical models of thrombosis to the problem of arterial clot formation revealed that to achieve platelet aggregation at pathologically high shear rates, platelet adhesion rates or adhesion strength must be increased to account for the thrombogenic effect of vWF. Hosseinzadegan & Tafti defined the adhesion rate for platelets as a linear function of the local wall shear rate (25). Du et al. modeled vWF-mediated platelet binding as a 100-fold increase in the bond formation rate that is enabled above a threshold flow elongation rate. Yazdani et al. increased the platelet adhesion strength by up to three orders of magnitude as a function of the local shear rate (26). Shankar et al. correlated the enhancement of platelet adhesion rate to the local shear rate by defining three shear regimes (low, medium, high), with a 20-fold enhancement in adhesion above a shear rate of 8000 $s^{-1}$ (27). Our own group has previously attempted to account for the effect of vWF by increasing the shear activation function of flowing platelets above a certain shear stress threshold (28).

While these models incorporate platelet-vWF interactions, they do not explicitly include vWF transport and flow-induced unfolding of vWF. Moreover, they correlate platelet adhesion parameters to local shear rate or elongation rate, but not both. vWF, on the other hand, exhibits distinctly different responses in simple shear and extensional flow (6–8, 29–31). In extensional flow such as planar or uniaxial extension, vWF multimers undergo an abrupt steady-state unfolding above a critical strain rate. In simple shear, however, above a critical shear rate, multimers undergo a tumbling motion where they extend and collapse periodically. Furthermore, vWF unfolding threshold for extensional flow was estimated to be two orders of magnitude lower than that for simple shear (8). Finally, this conformational change is reversible and involves hysteresis effects. This intricate behavior necessitates a more comprehensive modeling of vWF's response in complex flows and its subsequent involvement in thrombosis.

Our group recently presented a continuum model for the unfolding of vWF, where we represented the flow-induced conformational change of vWF from collapsed to stretched state using convection-diffusion-reaction (CDR) equations (32). The current work incorporates the vWF component into the multi-constituent model of thrombosis of Wu et al. (33, 34). The existing thrombosis model had been validated for low to moderate shear rates (100 – 1000 $s^{-1}$) but lacked the platelet-vWF interactions to predict thrombosis in high-shear environments. Here, we simulate vWF transport and conformational response in flow and model its thrombogenic effect, whereby the local concentration of stretched vWF amplifies the deposition rate of free-flowing platelets and reduces the shear cleaning of deposited platelets. We validate the model using two in vitro experiments representing thrombosis in diseased arteries and a clinical blood test used for screening for von Willebrand Disease.

## MATERIALS AND METHODS
### Multi-constituent model of thrombosis
*Equations of motion for blood*
Blood is treated as a multi-constituent mixture comprised of a fluid component, is assumed to behave as a linear viscous (Newtonian) fluid, and a thrombus component, treated as a porous medium. The fluid component, which accounts for the red blood cells (RBCs) and plasma, is governed by the continuity equation and the conservation linear momentum that includes interaction with the thrombus component:

$$\nabla \cdot \boldsymbol{v} = 0 \tag{1}$$

$$\rho \frac{D\boldsymbol{v}}{Dt} = -\nabla p + \mu \nabla^2 \boldsymbol{v} - \frac{C_2}{(1-\phi)} f(\phi)(\boldsymbol{v} - \boldsymbol{v_T}) \tag{2}$$

where $\boldsymbol{v}$ and $\boldsymbol{v_T}$ are the velocity of the fluid and thrombus in the inertial frame of reference, respectively. Since the thrombus is stationary relative to its substrate, $\boldsymbol{v_T}$ represents the velocity of the surface on which it forms; in the simulations reported in this study, $\boldsymbol{v_T} = 0$. $p$ in Eq. (2) is the pressure, $\rho$ is the density (1050 kg m$^{-3}$) and $\mu$ is the asymptotic viscosity (0.0035 Pa s) of blood. The scalar field $\phi$ represents the local volume fraction of the deposited platelets (thrombus). Free-flowing platelets are treated as scalar species that do not affect the physics of the flow. A growing thrombus, however, alters the local hemodynamics by means of a resistance force on the fluid. This heuristic approach is similar to one adopted by Leiderman and Fogelson (35). The resistance term is given by the last term on the right hand side of Eq. (2), in which $f(\phi) = \phi(1 + 6.5\phi)$ is the hindrance function from Wu et al. (33), who in turn adapted the model of Johnson et al. (36), based on the assumption that platelet aggregates behave like particulate sediment. The coefficient $C_2 = 2\times10^6$ kg m$^{-3}$ s$^{-1}$ is a constant selected to match the permeability of platelet-rich white clots formed under high-shear conditions, reported by Du et al (37). (The simulation method and results for permeability calculations are available in the Supporting Material.) To avoid singularity, the denominator $(1 - \phi) \in [\varepsilon, 1]$, where $\varepsilon$ is a small number.

Equations (1) and (2) were solved using OpenFOAM's *pimpleFoam* transient solver for incompressible Newtonian fluids that uses the PIMPLE (merged PISO-SIMPLE) algorithm (38, 39). The thrombus resistance term in Equation (2) was discretized as an implicit source term using OpenFOAM's *fvm::Sp* function (40). The scalar shear rate $\dot{\gamma}$ used throughout the thrombosis model was computed as:

$$\dot{\gamma} = \sqrt{2Tr(\boldsymbol{D} \cdot \boldsymbol{D})} \tag{3}$$

where $\boldsymbol{D}$ is the symmetric part of the velocity gradient tensor, $\boldsymbol{D} = {1}/{2}\,[\nabla \boldsymbol{v} + (\Delta \boldsymbol{v})^T]$.

*Convection-diffusion-reaction equations for thrombosis*

The thrombosis model includes six fundamental mechanisms involved in coagulation: (1) platelet activation, (2) platelet deposition, (3) thrombus propagation, (4) thrombus erosion by shear cleaning, (5) thrombus inhibition, and (6) thrombus-fluid interaction. The model considers four categories (states) of platelets, five biochemical species, and two vWF species – summarized in Table 1.

Table 1. Categories of platelet, biochemical, and vWF species in the model

|  | **Model abbreviation, [$C_i$]** | **Description** |
|---|---|---|
| Platelet (PLT) species | RP | Resting (unactivated) PLTs |
|  | AP | Activated PLTs |
|  | $RP_d$ | Deposited Resting PLTs |
|  | $AP_d$ | Deposited Activated PLTs |
| Biochemical species | $a_{pr}$ | PLT-released agonists (ADP) |
|  | $a_{ps}$ | PLT-synthesized agonists (TxA$_2$) |
|  | TB | Thrombin |
|  | PT | Prothrombin |
|  | AT | Antithrombin (ATIII) |
| vWF species | vWF$_c$ | vWF in a collapsed conformation |
|  | vWF$_s$ | vWF in a stretched conformation |

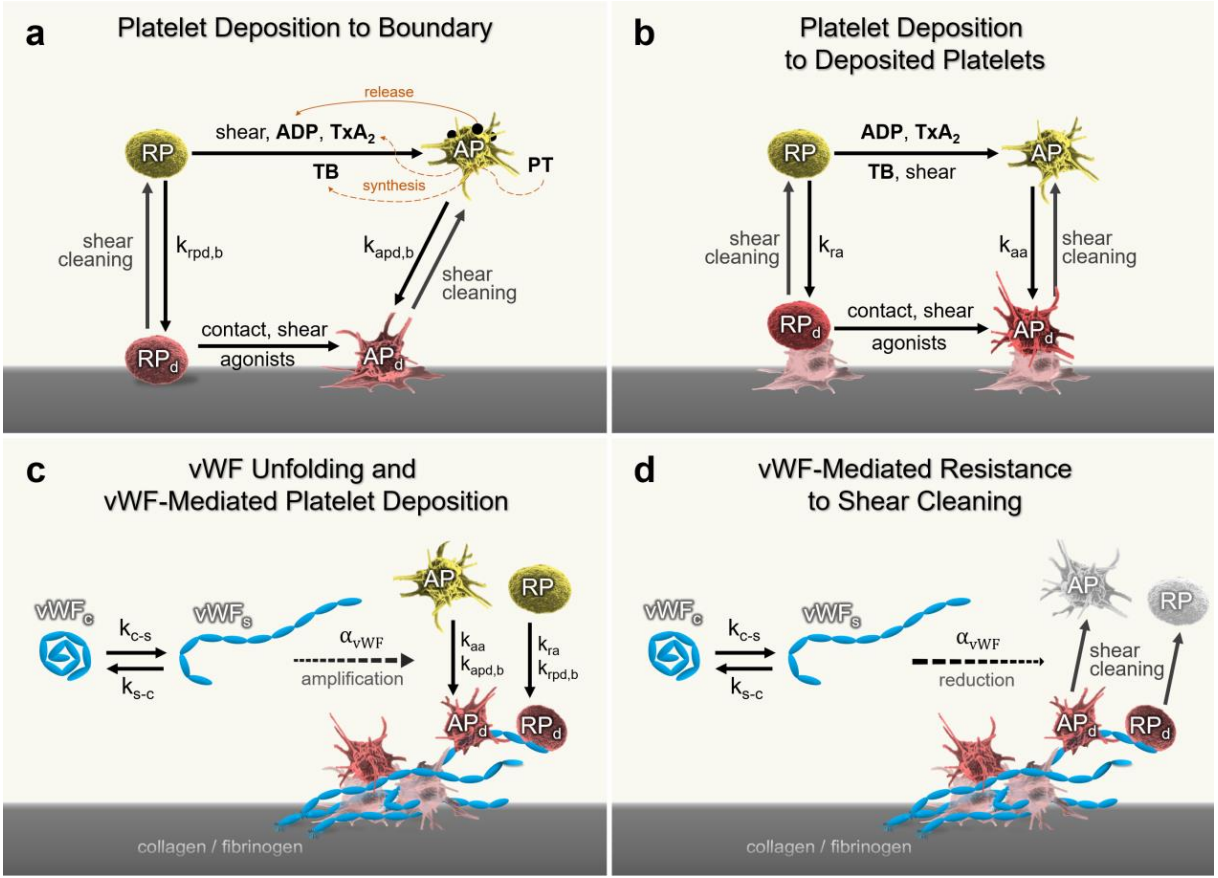

Fig. 1. Schematic depiction of the thrombosis model, showing (*a-b*) platelet activation by agonists and shear stress, platelet deposition, aggregation, and shear cleaning. The constants $k_{rpd,b}$, $k_{apd,b}$, $k_{ra}$, $k_{aa}$ refer to the reaction rates for inter-conversion of the associated platelet states (deposition), where the suffix *b* refers to the reaction with the boundary. (*c-d*) vWF unfolding in flow leads to concentration-dependent increase in platelet deposition and aggregation rates, as well as reduction in shear cleaning rates of deposited platelets.

The local concentration of resting (unactivated) platelets in the flow field, [RP], can be activated and converted into [AP] via two mechanisms. Biochemical activation occurs when the aggregate concentration of agonists [$a_{pr}$] (ADP), [$a_{ps}$] (TxA$_2$), and [TB] (thrombin) exceeds a critical value. The criterion for mechanical shear activation is adopted from Goodman et al. (41) and Hellums (42). Following the approach of Folie & McIntire and Hubbell & McIntire (43, 44), upon activation, platelets release ADP and continually synthesize TxA$_2$ in the activated state. Thrombin is synthesized on the activated platelet phospholipid membrane from prothrombin, [PT], and is inhibited by anti-thrombin III, [AT], mediated by heparin. (See Fig. 1 *A*.)

Free-flowing platelets can deposit onto boundaries (surfaces) or on previously deposited platelets – [RP$_d$] and [AP$_d$] – constituting a platelet clot (Fig. 1 *B*). Activated platelets [AP] deposit at a greater rate than [RP]. For both states, the rates of deposition are substrate-specific. Shear stress exerted by the fluid component can clear deposited platelets from surfaces and erode the clot periphery. Resting deposited platelets, [RP$_d$], are cleared at a greater rate than activated deposited platelets, [AP$_d$].

Two vWF species correspond to the two conformational states of vWF multimers: collapsed, [vWF$_c$], and stretched, [vWF$_s$]. The two states are interconvertible such that the total vWF concentration is conserved. The presence of stretched [vWF$_s$] multimers has a local thrombogenic effect via two mechanisms: it amplifies the deposition rates of free-flowing platelets (Fig. 1 *C*) and reduces the shear cleaning of deposited platelets (Fig. 1 *D*).

The transport of the species in the flow field is described by a set of convection-diffusion-reaction equations:

$$\frac{\partial [C_i]}{\partial t} + div(\boldsymbol{v}[C_i]) = div(D_i \nabla [C_i]) + S_i \tag{4}$$

where $[C_i]$ is the concentration of species *i*; $D_i$ is the diffusivity of species *i* in blood; and $S_i$ is the reaction source term for species *i*. Note that the subscript *i* represents the species listed in Table 1. The deposition/cleaning of platelets is governed by concentration rate equations, which are simplified from Eq. (4) in the absence of convective and diffusive terms for [RP$_d$] and [AP$_d$]:

$$\frac{\partial [C_i]}{\partial t} = S_i \tag{5}$$

The local volume fraction of the deposited platelets (thrombus), $\phi$, is evaluated as:

$$\phi = \begin{cases} \dfrac{[RP_d] + [AP_d]}{PLT_{max}} & \text{in space (internal field)} \\ \dfrac{[RP_d] + [AP_d]}{PLT_{s,max}} & \text{on surfaces (boundary field)} \end{cases} \tag{6}$$

where $PLT_{s,max} = 7 \times 10^{10} PLT\ m^{-2}$ is the total capacity of the surface for platelets, $PLT_{max} = \frac{PLT_{s,max}}{Dia_{PLT}}$ is the maximum concentration of platelets in space, and $Dia_{PLT} = 2.78 \times 10^{-6} m$ is the hydraulic diameter of platelets.

Reactions at a boundary, including deposition of platelets to surfaces, are modeled by surface-flux boundary conditions (BCs), following the approach of Sorensen et al. (45). Further detailed mathematical description of the thrombosis model, including the specific values and expressions for all the source terms and their parameters, can be found in in the Supporting Material, as well as publication of Wu et al. (33), with additional validation in Zhussupbekov et al. (34).

## vWF unfolding and vWF-mediated platelet deposition

This work integrates our recently introduced continuum model of vWF unfolding (32) into previously published multi-constituent model of thrombosis of Wu et al. (33). In all simulations, vWF is initially present in the collapsed conformation as [vWF$_c$]. The stretched [vWF$_s$] is produced by means of [vWF$_c$] unfolding. The transport and interconversion of the vWF species are described by CDR equations:

$$\frac{d}{dt}[vWF_c] + div(\boldsymbol{v}[vWF_c]) = div(D_{vWF} \nabla [vWF_c]) - k_{c\text{-}s}[vWF_c] + k_{s\text{-}c}[vWF_s] \tag{7}$$

$$\frac{d}{dt}[vWF_s] + div(\boldsymbol{v}[vWF_s]) = div(D_{vWF} \nabla [vWF_s]) + k_{c\text{-}s}[vWF_c] - k_{s\text{-}c}[vWF_s] \tag{8}$$

where $D_{vWF} = 3.19 \times 10^{-11}\ m^2 s^{-1}$ is the diffusivity of vWF species (46) and $k_{c\text{-}s}$ and $k_{s\text{-}c}$ are the collapsed-to-stretched and stretched-to-collapsed conversion rates, respectively.

Dilute flexible polymers such as vWF demonstrate distinct conformational responses to different types of flow (6–8, 29–31). In extensional flows such as planar or uniaxial extension, polymers undergo an abrupt steady-state unfolding above a critical strain rate. In simple shear, however, polymers undergo a tumbling motion where they extend and collapse periodically above a critical shear rate. Babcock et al (47) studied the behavior of polymers near the critical point of coil-stretch transition between simple shear and extensional flow. They classified the flow according to its extensional or rotational quality using the flowtype parameter, $\lambda$:

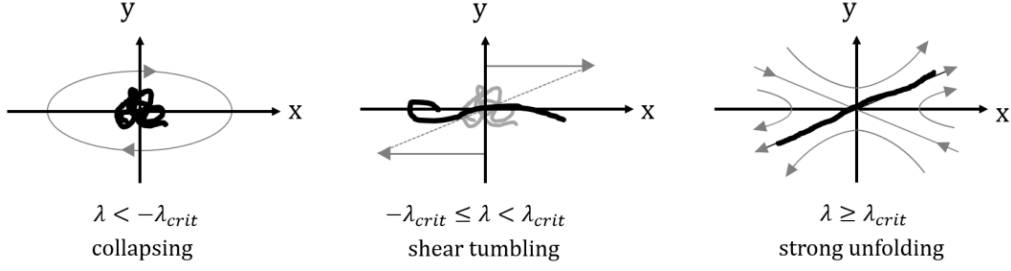

Fig. 2. vWF chain configuration in flows with various flowtype, $\lambda$ = [-1, 1], where $\lambda_{crit}$ = 0.0048 (47). When $\lambda < -\lambda_{crit}$, the chain is collapsed. In near-shear flows, $\lambda \approx 0$, the chain undergoes a tumbling motion (unfolding-collapsing cycles). When $\lambda \geq \lambda_{crit}$, the chain is fully extended with minimal chain fluctuations. Adapted from Woo & Shaqfeh (48), with the permission of AIP Publishing.

$$\lambda = \frac{\|\mathbf{D}\| - \|\mathbf{\Omega}\|}{\|\mathbf{D}\| + \|\mathbf{\Omega}\|} \tag{9}$$

where $\mathbf{D}$ and $\mathbf{\Omega}$ are the symmetric and antisymmetric parts of the velocity gradient tensor, respectively, so that $\nabla \mathbf{v} = \mathbf{D} + \mathbf{\Omega}$, where $\mathbf{D} = \frac{1}{2}(\nabla \mathbf{v} + (\Delta \mathbf{v})^T)$ and $\mathbf{\Omega} = \frac{1}{2}(\nabla \mathbf{v} - (\Delta \mathbf{v})^T)$. Their magnitudes are given by $\|\mathbf{D}\| = \sqrt{\frac{1}{2}\mathbf{D}:\mathbf{D}}$ and $\|\mathbf{\Omega}\| = \sqrt{\frac{1}{2}\mathbf{\Omega}:\mathbf{\Omega}}$, where (:) is the double inner product.

Then, $\lambda$ = 1 in purely extensional flow, $\lambda$ = -1 in pure rotation, and $\lambda$ = 0 in simple shear.

Fig. 2 illustrates the three flow regimes based on $\lambda$. vWF will experience strong unfolding in flows with $\lambda \geq \lambda_{crit}$, shear tumbling in -$\lambda_{crit} \leq \lambda < \lambda_{crit}$, and remain collapsed if $\lambda < -\lambda_{crit}$. The critical value for flowtype, $\lambda_{crit}$ = 0.0048, is adopted from the experiment of Babcock *et al.* These three regimes are modeled through simultaneous action of the stretched-to-collapsed and collapsed-to-stretched conversion rates, $k_{s-c}$ and $k_{c-s}$, respectively, as described below.

*Shear tumbling*
In simple shear and near-shear flows, -$\lambda_{crit} \leq \lambda < \lambda_{crit}$, the stretched-to-collapsed conversion rate $k_{s-c}$ is constant, meaning the default behavior of vWF chains is to collapse into a globular state. The collapsed-to-stretched conversion rate $k_{c-s}$, on the other hand, is a function of shear rate, with its asymptotic value matching $k_{s-c}$. Thus, at high shear rates, this results in an equal steady-state distribution of [vWF$_c$] and [vWF$_s$], reflecting the approximation that when all vWF chains are undergoing shear tumbling, at any given moment roughly half of the total vWF concentration would be in a stretched conformation and the other half in a collapsed conformation. This mirrors the experimental and simulation results where the ensemble-averaged fractional extension of vWF approached 0.5 with increasing shear rate (7, 31, 49).

The unfolding rate in simple shear, $k_{c-s}^{shear}$, follows a function proposed by Lippok et al. for shear-dependent cleavage rate of vWF (11). Since this function describes the availability of vWF monomers for enzymatic cleavage the unfolding rate is assumed to follow the same shape, given by Eq. (10):

$$k_{c-s}^{shear}(\dot{\gamma}) = \frac{k'}{1 + \exp\left(-\frac{\dot{\gamma} - \dot{\gamma}_{1/2}}{\Delta \dot{\gamma}}\right)} \tag{10}$$

$$k_{s-c}^{shear} = k' \tag{11}$$

where $\dot{\gamma}_{1/2}$ = 5522 s$^{-1}$ is the half-maximum shear rate for unfolding and $\Delta \dot{\gamma}$ = 1271 s$^{-1}$ is the width of the transition from Lippok et al.; $k'$ is the nominal state conversion rate, $k' = \frac{1}{t_{vWF}}$, where $t_{vWF}$ = 50 ms is the vWF unfolding time (9, 50, 51). Although $t_{vWF}$ is used as a constant, the actual unfolding rate $k_{c\text{-}s}^{shear}$ is a function of shear rate $\dot{\gamma}$ as shown in Fig. 3. The collapsing rate $k_{s\text{-}c}^{shear}$ is set to $k'$ as shown in Eq. (11).

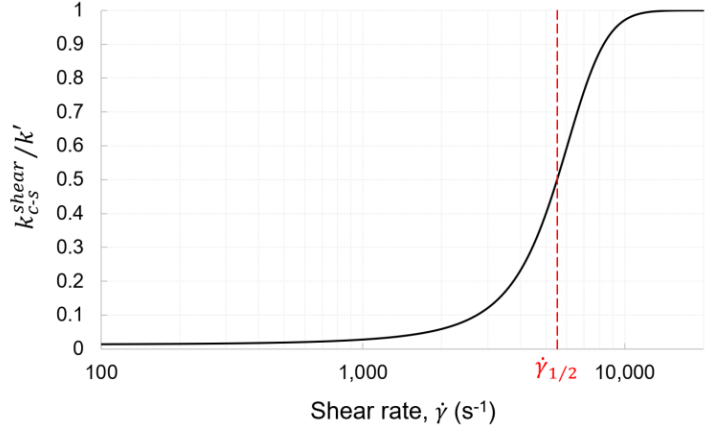

Fig. 3. Plot of the vWF unfolding rate in simple shear, $k_{c\text{-}s}^{shear}$ – corresponding to Eq. (10) – normalized by the nominal state conversion rate, $k'$, as a function of shear rate, $\dot{\gamma}$. The half-maximum shear rate $\dot{\gamma}_{1/2}= 5522$ s$^{-1}$ is the value around which vWF has been reported to unfold in simple shear (7, 11, 49).

*Strong unfolding*
Babcock et al. found that in extension-dominated flows with flowtype values $\lambda \geq \lambda_{\text{crit}}$, the average stretch of the polymer is determined solely by the modified Weissenberg number:

$$Wi_{eff} = Wi\sqrt{\lambda} = (\|\boldsymbol{D}\| + \|\boldsymbol{\Omega}\|)\tau_{rel}\sqrt{\lambda} \tag{12}$$

where $\tau_{rel}$ is the polymer relaxation time. Therefore, in the strong unfolding regime, vWF unravels if $Wi_{eff}$ exceeds the critical value $Wi_{eff,crit}$ and the unfolding rate $k_{c\text{-}s}^{strong}$ scales with $Wi_{eff}$ according to Eq. (13), while $k_{s\text{-}c}^{strong}$ is zero.

If $Wi_{eff}$ falls below $Wi_{eff,crit}$, the stretched [vWF$_s$] will experience hysteresis (30, 52) and will not collapse back until $Wi_{eff} < Wi_{eff,hyst}$, as shown in Eq. (14).

$$k_{c\text{-}s}^{strong} = \begin{cases} k'\dfrac{Wi_{eff}}{Wi_{eff,crit}}, & Wi_{eff} \geq Wi_{eff,crit} \\ k_{c\text{-}s}^{shear}, & Wi_{eff} < Wi_{eff,crit} \end{cases} \tag{13}$$

$$k_{s\text{-}c}^{strong} = \begin{cases} 0, & Wi_{eff} \geq Wi_{eff,hyst} \\ k', & Wi_{eff} < Wi_{eff,hyst} \end{cases} \tag{14}$$

The unfolding threshold and the hysteresis value are adopted from Brownian dynamics simulations of Sing and Alexander-Katz (30), where $Wi_{eff,crit} = 0.316$ and $Wi_{eff,hyst} = 0.053$, with additional mathematical details available in the Supporting Material.

*Collapsed conformation*
In flows with a dominant rotational component, $\lambda < -\lambda_{\text{crit}}$, [vWF$_c$] remains collapsed and [vWF$_s$] reverts to a globular state, so $k_{c\text{-}s} = 0$ and $k_{s\text{-}c} = k'$. This, combined with the defined thresholds for unfolding and collapsing in the two other regimes, localizes the action of the vWF to sites of pathological hemodynamic conditions.

*Thrombogenic effect of vWF*

The presence of vWF in stretched conformation has a twofold local thrombogenic effect: it amplifies the deposition rates of free-flowing platelets, $k_m$, (Fig. 1 *C*) and reduces the shear cleaning of deposited platelets, $\tau_m$, (Fig. 1 *D*) as a function of the local [vWF$_s$]:

$$\alpha_{vWF} = \begin{cases} \dfrac{[\text{vWF}_s]}{[\text{vWF}_s]_{crit}}, & [\text{vWF}_s] \geq [\text{vWF}_s]_{crit} \\ 1, & [\text{vWF}_s] < [\text{vWF}_s]_{crit} \end{cases} \quad (15)$$

$$k_m^{vWF} = \alpha_{vWF} k_m \quad (16)$$

$$\tau_m^{vWF} = \alpha_{vWF} \tau_m \quad (17)$$

where $k_m$ represents the deposition rate constants $k_{ra}$, $k_{aa}$, $k_{rpd,b}$, $k_{apd,b}$, and $\tau_m$ represents the shear cleaning rate constants $\tau_{rd}$, $\tau_{ad}$, $\tau_{rd,b}$, $\tau_{ad,b}$. (See Table S6 for description of terms.) The critical concentration of stretched vWF is set as $[\text{vWF}_s]_{crit} = 0.01([\text{vWF}_c] + [\text{vWF}_s])$ such that $\alpha_{vWF}$ is bound between 1 and 100.

## Simulation of in vitro thrombosis

To validate the model of vWF-mediated thrombosis, three experiments were simulated: thrombotic occlusion of a stenotic tube, the PFA-100® clinical blood test, and thrombus formation in a stagnation point flow microchannel. These validation cases feature hemodynamics of high shear rates and extensional flow gradients required for vWF unfolding, as well as a collagen substrate for initiating thrombus formation. We have previously calibrated the substrate-specific platelet deposition rates for collagen (34), and we use the same baseline values (Table 2) in present simulations.

The thrombosis model was numerically implemented in an open-source finite volume software library OpenFOAM (The OpenFOAM Foundation Ltd). The simulations were performed using OpenFOAM v6 and post-processing was done in ParaView 5.6 (53).

In all cases, prior to launching the thrombosis simulation, a steady-state flow solution was obtained first to improve stability at the start of the thrombosis run. The convergence criteria for the residual errors in transient simulations were set to 1×10$^{-4}$, with second-order discretization schemes. A dual time step strategy was employed to reduce the computational cost. A time step of $\Delta t_{CDR}$ = 1×10$^{-2}$ s was used to solve the species equations and thrombus evolution, while the flow equations were solved with a time step of $t_{CFD} = \dfrac{t_{CDR}}{r_{flow}}$ = 1×10$^{-8}$ s, ensuring the Courant–Friedrichs–Levy stability condition, CFL ≤ 1. This approach has previously been used in Méndez Rojano et al. (54), where the thrombosis simulation results were unaffected by different values of the scaling factor $r_{flow}$ spanning four orders of magnitude. Additionally, similar strategies have been employed in multi-scale models of platelet activation (55).

Table 2. Collagen-specific platelet deposition and shear cleaning constants at the boundary.

| Term | Value | Units | Description |
| --- | --- | --- | --- |
| $k_{apd,b}$ | $3.0 \times 10^{-6}$ | m s$^{-1}$ | [AP]-to-boundary deposition rate |
| $k_{rpd,b}$ | $1.0 \times 10^{-8}$ | m s$^{-1}$ | [RP]-to-boundary deposition rate |
| $\tau_{ad,b}$ | 0.1 | dyne cm$^{-2}$ | Boundary shear cleaning related constant for [AP$_d$] in the expression $\left(1 - exp(-0.0095\, \tau/\tau_{ad,b})\right)$ from Goodman et al. (41) |
| $\tau_{rd,b}$ | 0.05 | dyne cm$^{-2}$ | Boundary shear cleaning related constant for [RP$_d$] in the expression $\left(1 - exp(-0.0095\, \tau/\tau_{rd,b})\right)$ from Goodman et al. (The smaller value for [RP$_d$] translates to greater cleaning compared to [AP$_d$].) |

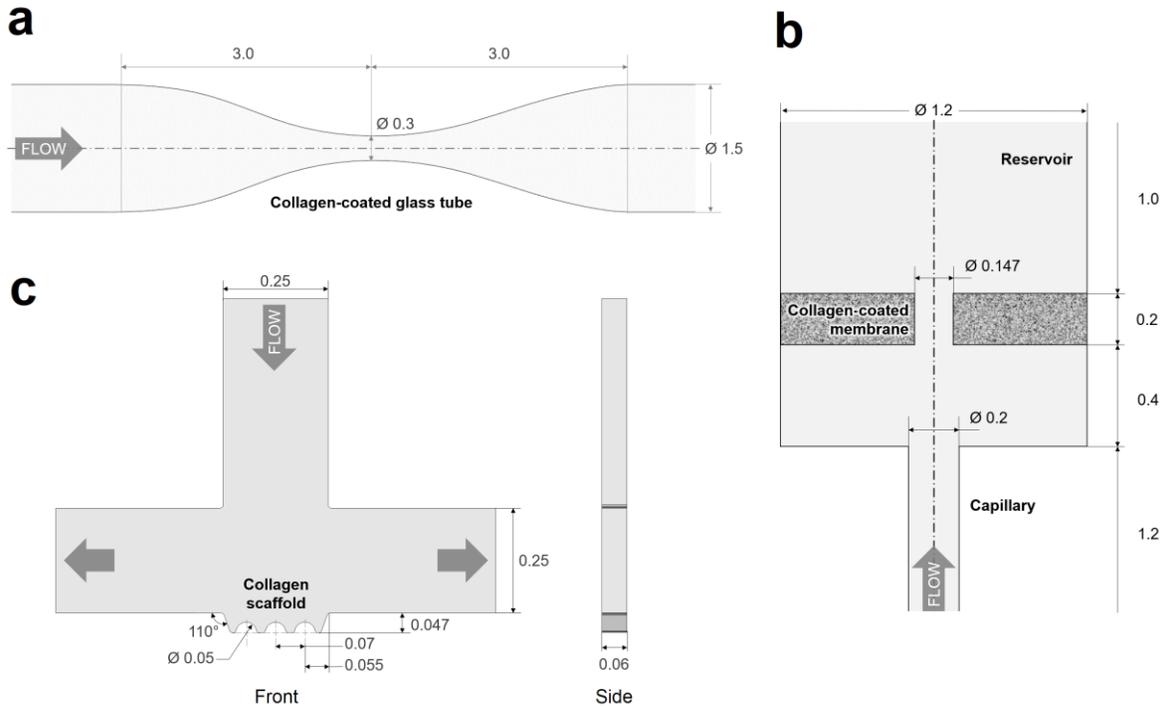

Fig. 4. Schematics of the simulated in vitro systems. Length dimensions are in millimeters. (a) Collagen-coated tube with a stenotic section mimicking an atherosclerotic artery (56). (b) PFA-100® circular cartridge where blood is aspirated through a central aperture of a collagen-coated membrane. (c) Stagnation point flow channel where blood is perfused directly perpendicular to a collagen scaffold formed over cylindrical micro-posts (57).

*Stenotic tube*

Ku and colleagues studied thrombosis under arterial flow conditions employing a collagen-coated tube with a stenotic test section that mimics the shape of an atherosclerotic plaque, creating hemodynamic conditions of great shear rate gradients, extensional flow, high shear rates, and short exposure times (56, 58). This experimental setup produced thrombotic occlusion at the throat of the stenosis within 4-8 minutes of whole blood perfusion.

Flow in an axisymmetric domain with 80% stenosis was simulated where the initial lumen diameter of 1.5 mm contracts to 0.3 mm at the throat of the stenosis. (See Fig. 4 *a*.) The flow rate was fixed to generate the wall shear rate of 5000 s$^{-1}$ at the narrowest section of the tube, prescribed as a developed velocity profile at the inlet. Following the collagen coating protocol used in the experiment, the reactive BCs for platelet deposition were applied at the walls starting 1 mm upstream of the stenotic section (4 mm upstream of the apex of the stenosis) and extending to the outlet of the domain. Platelets were introduced at the inlet at $3\times10^{14}$ PLT m$^{-3}$ concentration with 1% background activation level, [RP] = $2.97\times10^{14}$ PLT m$^{-3}$ and [AP] = $3\times10^{12}$ PLT m$^{-3}$. Heparin concentration matched the experimental protocol at 3.5 U ml$^{-1}$ ($7.3\times10^5$ nmol m$^{-3}$ using specific activity of 300 U mg$^{-1}$ and molecular weight of 16 kDa (59, 60)). vWF was introduced entirely in a collapsed conformation, [vWF$_c$] = 1000 nmol m$^{-3}$ and [vWF$_s$] = 0. Inlet BCs for the rest of the species followed (33, 45) and are summarized in Table 3.

A 2-D axisymmetric mesh was composed of 80,000 hexahedral cells. Mesh independence study was conducted where the cell size was refined by a factor of 2 in every dimension to generate three levels of grid (coarse, medium, fine). Shear rate near the stenosis apex was assessed to verify that the relative error between the medium and fine grids was below 1%. Details of the mesh independence study for this and following cases are presented in Table 4.

Table 3. Inlet concentration of species prescribed as Dirichlet boundary conditions.

| Species | Inlet concentration | |
|---|---|---|
| [RP] | $2.97\times10^{14}$ PLT m$^{-3}$ | Total count |
| [AP] | $3\times10^{12}$ PLT m$^{-3}$ | $3\times10^{14}$ PLT m$^{-3}$ |
| [a$_{pr}$] (ADP) | 0 nmol m$^{-3}$ | |
| [a$_{ps}$] (TxA$_2$) | 0 nmol m$^{-3}$ | |
| [TB] | 0 nmol m$^{-3}$ | |
| [PT] | $1.1\times10^6$ nmol m$^{-3}$ | |
| [AT] | $2.844\times10^6$ nmol m$^{-3}$ | |
| [vWF$_c$] | 1000 nmol m$^{-3}$ | |
| [vWF$_s$] | 0 nmol m$^{-3}$ | |

Table 4. Mesh independence study of the three simulation cases. Relative error in computed shear rate values is reported for medium and fine grid levels.

| | Problem dimension | Grid refinement factor | Coarse grid cell count | Medium grid cell count | Fine grid cell count | Relative error, % |
|---|---|---|---|---|---|---|
| **Stenotic tube** | 2-D axisymmetric | 2 | 20,000 | 80,000 | 320,000 | 0.06 |
| **PFA-100®** | 2-D axisymmetric | 2 | 17,455 | 68,650 | 274,600 | 0.76 |
| **Stagnation channel** | 3-D, quarter-symmetry | 1.5 | 409,937 | 1,007,125 | 3,116,822 | 0.14 |

*Platelet function analyzer PFA-100®*

The platelet function analyzer PFA-100® (Siemens Erlangen, Germany) is an in vitro system used as a screening test for patients with impaired primary hemostasis (61). PFA-100® is highly sensitive to the influence of plasma vWF and is routinely used for screening for von Willebrand disease (62–64). Fig. 4 *b* shows a cross-section sketch of the PFA-100® test cartridge whose main component is a collagen-coated membrane with a 140-μm central aperture. The membrane is also coated with either ADP or epinephrine to promote platelet activation. Whole blood is aspirated through a capillary towards the membrane (bottom to top), and as blood flows through the central aperture of the membrane, a clot forms until occlusion is achieved. The normal reference range for closure time of the collagen/ADP test cartridges is 60-120 s (65, 66).

Flow was simulated in an axisymmetric domain representing the PFA-100® system (Fig. 4 *b*). A developed velocity profile with the mean velocity of 0.0637 m s$^{-1}$ was prescribed at the inlet to achieve the design-specified peak shear rate of 6000 s$^{-1}$ in the membrane aperture (64). Reactive BCs for platelet deposition were applied at the walls of the membrane and a diffusive flux of ADP = $1\times10^8$ nmol m$^{-2}$ s$^{-1}$ was prescribed to mimic the shedding of ADP from the membrane. (Estimation of the flux value and its effect on the thrombus growth rate are provided in the Supporting Material.) Inlet BCs for the biochemical species were identical to the previous case and are summarized in Table 3.

*Stagnation point flow channel*

Stagnation point flow is one of the classical examples of flows with strong extensional deformations (67, 68), pertinent to vWF unfolding. Herbig and Diamond studied thrombus formation in a stagnation point flow by perfusing whole blood directly perpendicular to a collagen thrombotic surface (57). They employed a T-junction channel comprised of a top inlet and two side outlets (Fig. 4 *c*). Prior to blood perfusion, a collagen plug was formed over an array of cylindrical micro-posts at the stagnation point, which acted as a substrate for platelet adhesion. The experiment was performed at two flow rates that produced thrombi of distinct morphology: at 10 μL min$^{-1}$, clots grew into dendritic structures, while thrombi formed at 100 μL min$^{-1}$ appeared smoother and more connected. When platelet-vWF interactions were blocked, platelet aggregation was reduced in the low flow rate case and was practically eliminated in the high flow rate case.

The computational domain was truncated at the top row of the cylinders to mimic the collagen plug, as shown in Fig. 4 *c*. Reactive BCs for platelet deposition were applied on the side wall surfaces within the collagen scaffold region. Flow rates of 10 and 100 μL min$^{-1}$ were imposed at the inlet as a uniform velocity profile. Inlet BCs for the biochemical species are summarized in Table 3. Taking advantage of the symmetry of the channel, a quarter of the geometry was simulated to reduce the computational cost. A semi-structured hex-dominant mesh of 1 million cells was generated using OpenFOAM's snappyHexMesh utility, and increased mesh refinement was applied around the scaffold region.

## RESULTS
### Stenotic tube

Fig. 5 *a* demonstrates the time progression of simulated thrombus growth in the stenotic tube. While the entire lumen starting 4 mm upstream of the apex of the stenosis was set as reactive boundary, the platelet deposition was localized to the apex. Deposited platelets started to accumulate in the boundary field within the first 70 seconds of the simulation and subsequently propagated into the domain. As the thrombus grew, wall shear rate on the lumen of the tube increased by two orders of magnitude (Fig. 5 *b*). Despite the increased shear rate, rapid platelet accumulation continued up to total occlusion at around 5 minutes. Fig. 5 *c* displays the experimental time-lapse images of thrombus formation in a stenotic tube (58) at similar time points.

Fig. 6 shows the concentration of stretched vWF on a slice through the middle plane of the tube. [vWF$_c$] unfolds into [vWF$_s$] at the apex of the stenosis due to elevated shear rate near the walls and high extensional flow gradients, $Wi_{eff}$, in the contraction and expansion sections. The increased concentration of [vWF$_s$] amplifies the platelet deposition rates and reduces the shear cleaning rate of deposited platelets. A growing thrombus shrinks the lumen creating larger shear rates and extensional flow gradients, which forms a positive feedback loop with vWF unfolding.

The simulated thrombus growth in the stenotic tube followed the two distinct phases described by Ku and colleagues (5, 22, 56), as shown in Fig. 7 *a-b*. In the first phase, called the *lag time*, platelets start to accumulate around the apex of the stenosis, but the deposition is limited to the boundary field and most of the deposited platelets are in a resting (unactivated) state. (See Fig. S4.) Past the 1 minute mark, however, bulk activation of deposited platelets occurs, which triggers the *rapid platelet accumulation* (RPA) phase. In this phase, the thrombus grows into the throat at an average rate of $6.2 \times 10^{-4}$ mm$^3$ s$^{-1}$, which is comparable to the experimental growth rate of $5.7 \pm 2.7 \times 10^{-4}$ mm$^3$ s$^{-1}$ reported for the 6500 s$^{-1}$ case in (58).

Bark, Para, and Ku conducted a control experiment where blood was perfused through a straight collagen-coated tube (in the absence of stenosis) at a wall shear rate of 1000 s$^{-1}$, which produced only a thin layer of platelets after 12 minutes of perfusion (56). In a corresponding simulation, platelets deposited only in the first layer of mesh cells adjacent to the wall after 12 minutes of simulation time (Fig. S7), demonstrating the robustness of the model.

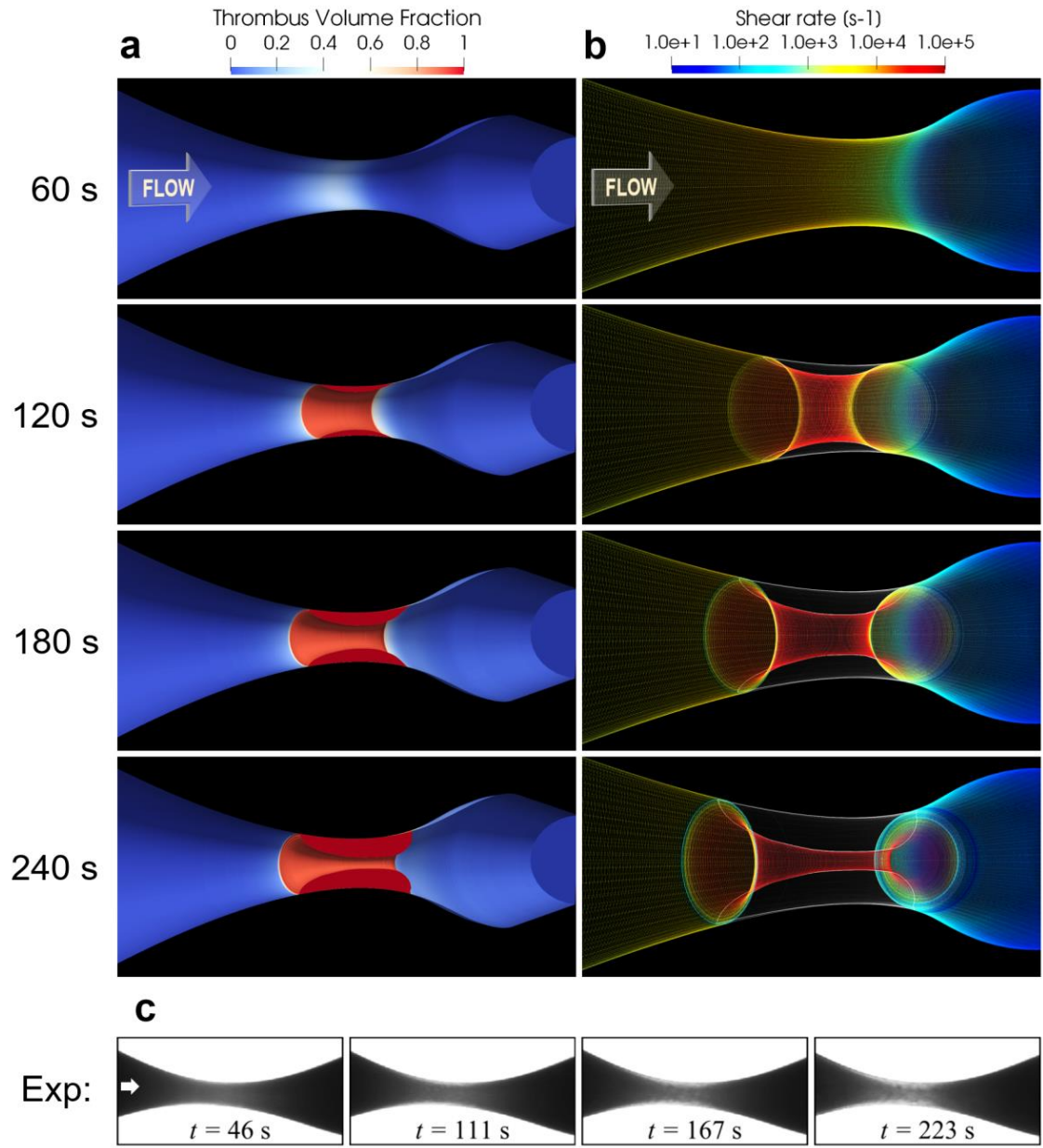

Fig. 5. Time progression of thrombotic occlusion in a stenotic tube. (a) 3-D rotational extrusion of the axisymmetric simulation showing the thrombus volume fraction. While the entire lumen of the tube was prescribed as a reactive boundary, thrombosis was localized to the apex of the stenosis. (b) Wall shear rate on the lumen using log scale. Peak shear rate at the apex of the stenosis increased by two orders of magnitude over the course of occlusion. (c) Experimental time-lapse images of thrombus formation in a stenotic tube from (58). Reprinted by permission from Elsevier, *Cardiovasc Eng Tech*, "High Shear Thrombus Formation under Pulsatile and Steady Flow", Casa & Ku, Copyright 2014 Springer Nature.

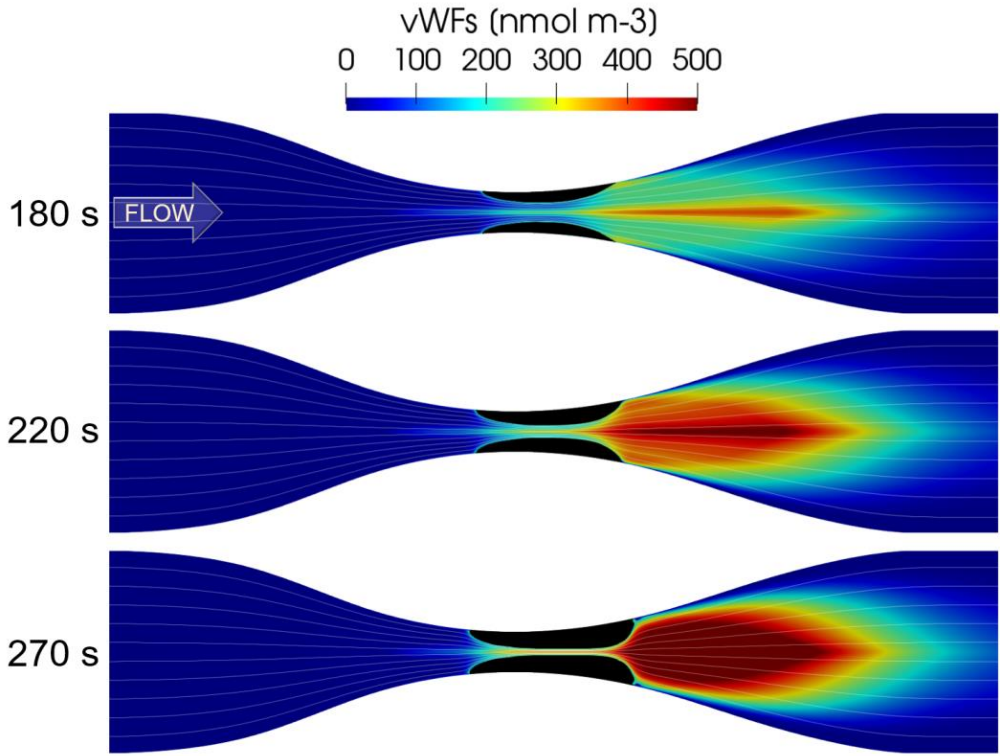

Fig. 6. Stretched vWF concentration, [vWF$_s$], shown on a slice through the meridional plane of the stenotic tube. vWF unfolding at the apex of the stenosis initiates the platelet aggregation. A growing thrombus (indicated in black) shrinks the lumen creating larger shear rates and extensional flow gradients, which forms a positive feedback loop with vWF unfolding.

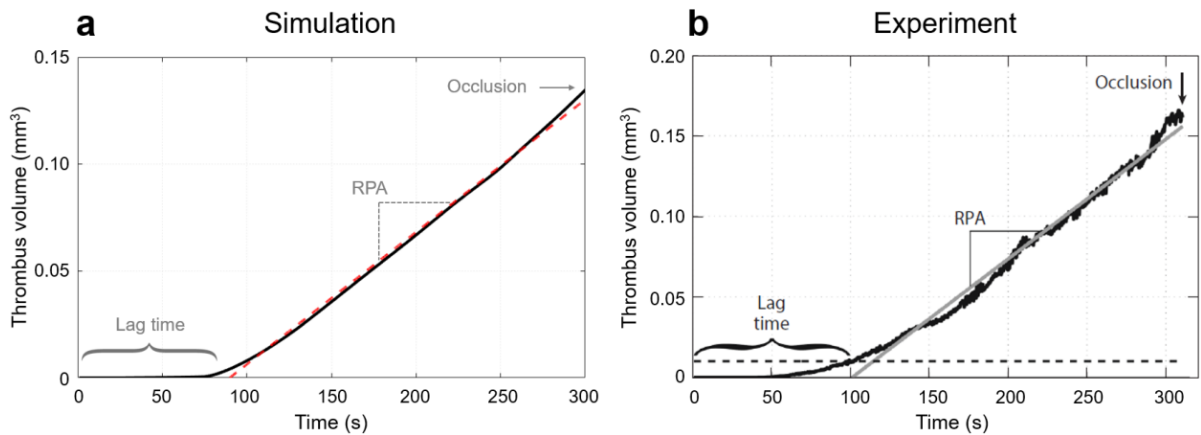

Fig. 7. Simulation reproduced the thrombus growth phases observed experimentally. (a) In the initial phase called the *lag time*, platelets accumulated in the boundary field and were mostly unactivated. Buildup of platelet agonists concentration and subsequent bulk activation of deposited platelets triggered the *rapid platelet accumulation* phase, culminating in lumen occlusion. (b) Experimental results from (5). Reprinted from *Journal of Vascular Surgery*, Casa & Ku, Vol 61, "Role of high shear rate in thrombosis", pp 1068-1080, Copyright (2015), with permission from Elsevier.

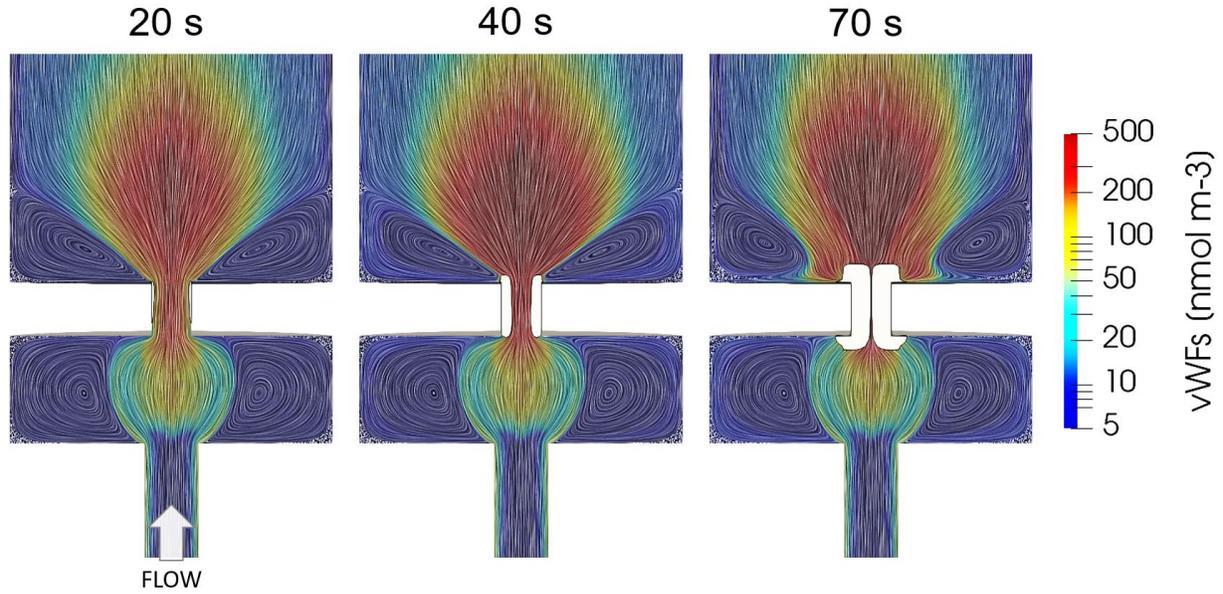

Fig. 8. Thrombotic occlusion of the membrane aperture in the PFA-100® device. Velocity LIC describes the flow field and color illustrates the stretched vWF concentration shown on the meridional plane. Thrombus is indicated in white.

## PFA-100®

Fig. 8 shows the time progression of the thrombotic occlusion in the PFA-100® device. Platelets deposited on the membrane walls within the aperture, and thrombus (indicated in white) started advancing within the first 20 seconds of simulation. The velocity line integral convolution (LIC) illustrates the flow field with several stages of contraction and expansion of streamlines, which creates strong vWF unfolding. (A detailed illustration of the unfolding rates can be found in Zhussupbekov et al. (32)) The increasing concentration of [vWF$_s$] sustains the platelet deposition resulting in occlusion of the aperture at around 70 seconds.

In addition to the baseline simulation, where inlet platelet concentration was $3\times10^{14}$ PLT m$^{-3}$ and [vWF$_c$] concentration was 1000 nmol m$^{-3}$, three clinical hemostatic deficiency conditions were simulated: thrombocytopenia, vWD Type 1, and vWD Type 3. Thrombocytopenia is a platelet deficiency where the platelet count falls below $1.5\times10^{14}$ PLT m$^{-3}$ (69). Therefore, we simulated thrombocytopenia by prescribing the platelet concentration at the inlet as 1/3 of the baseline case, at $1.0\times10^{14}$ PLT m$^{-3}$. vWD Type 1 is a partial quantitative deficiency in vWF (70), which was simulated by reducing the inlet [vWF$_c$] concentration by 50%. Finally, vWD Type 3 is a complete quantitative and functional deficiency, hence no vWF was introduced into the domain.

Table 5 shows simulation results compared to reported PFA-100® closure time for patients with normal platelet function and patients with hemostatic deficiencies (63). (Details of the statistical analysis of the reported data are presented in the Supporting Material.) Fig. 9 plots the percent occlusion, defined as percentage of the aperture diameter occupied by thrombus, and thrombus volume against time. Compared to the baseline case simulation, occlusion took more than twice as long for vWD Type 1 and three times longer for thrombocytopenia. vWD Type 3 case did not reach closure within 300 seconds of simulation, which is the maximum test time for the PFA-100®. The final thrombus volume at the moment of occlusion was similar in baseline and thrombocytopenia cases but was almost 2 times greater in the vWF Type 1 case (Fig. 9 *b*).

Table 5. Comparison of literature-derived and simulated closure time values (time to occlusion of the membrane aperture) for baseline and hemostatic deficiency cases

|  | Closure time, s | |
| --- | --- | --- |
|  | **Harrison 2002** | **Simulation** |
| **Baseline** | 90 ± 16 | 71 |
| **Thrombocytopenia** | 202 ± 89 | 212 |
| **vWD Type 1** | 161 ± 64 | 155 |
| **vWF Type 3** | > 300 | > 300 |

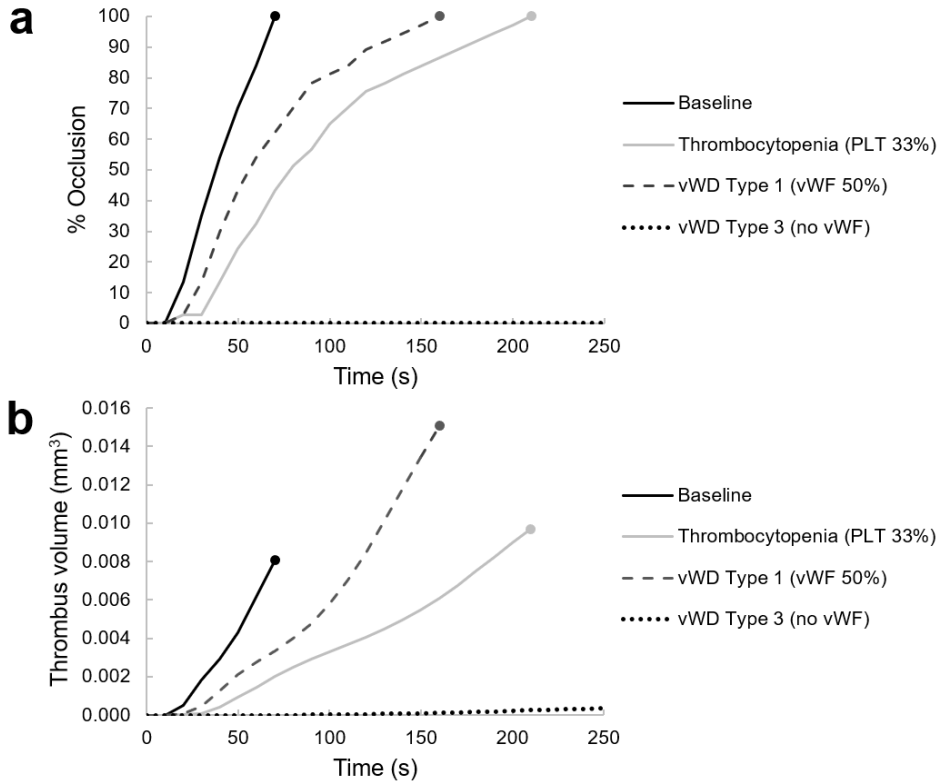

Fig. 9. PFA-100® simulation results for the baseline case and various hemostatic deficiencies. (a) Percentage of the aperture diameter occupied by thrombus plotted against simulation time. (b) Thrombus volume plotted against time. Solid circles indicate occlusion of the membrane aperture.

Fig. 10 highlights the difference in platelet deposition between the baseline and vWD Type 3 cases. At baseline levels of vWF, platelets deposited within the membrane aperture and quickly formed an occluding thrombus. In the absence of vWF, however, platelets could only aggregate in the low-shear stagnation zone at the flow reattachment point. (The flow structure that gives rise to this low-shear zone can be seen in Fig. 8.) Despite the same level of platelet count, no thrombus formed within the membrane aperture after 300 s of simulation in the vWD Type 3 case.

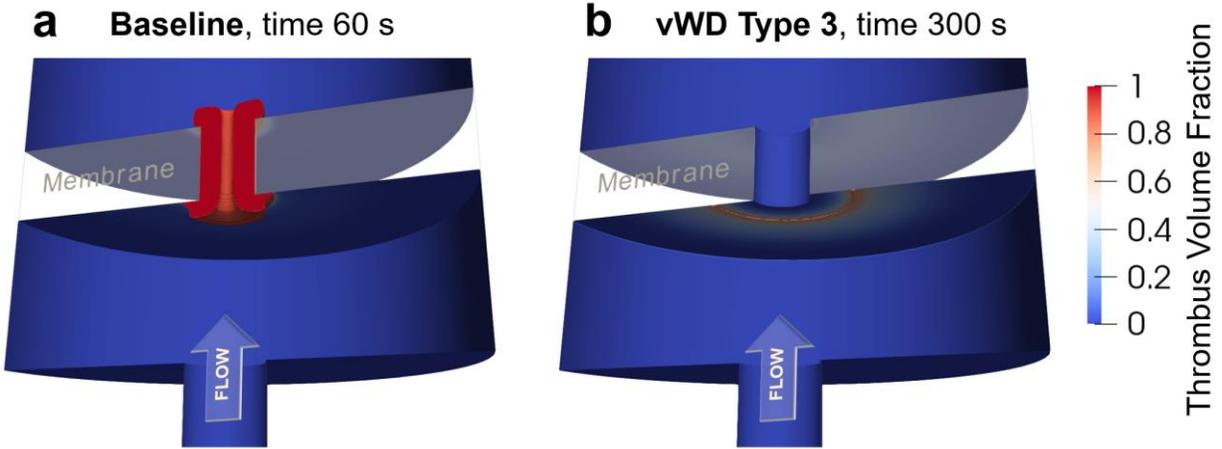

Fig. 10. Rotational extrusion of the PFA-100® computational domain and the thrombus. (a) At baseline level of vWF concentration, platelets deposited within the membrane aperture and quickly formed an occluding thrombus. (b) In the absence of vWF, platelets could only aggregate in the low-shear stagnation zone at the flow reattachment point.

### Stagnation point flow channel

Fig. 11 shows the velocity streamlines and wall shear rate in the stagnation point flow channel at two flow rates: 10 and 100 μL/min. An order of magnitude increase in flow rate generates a proportional increase in shear rate and $Wi_{eff}$ levels. Correspondingly, vWF unfolding was limited at 10 μL/min, whereas 100 μL/min produced extensive vWF unfolding, as shown in Fig. 12. The elevated wall shear rate in the inlet channel produced high [vWF$_s$] concentration near the walls, while extensional flow around the stagnation region induced strong unfolding close to the middle Z-plane of the channel.

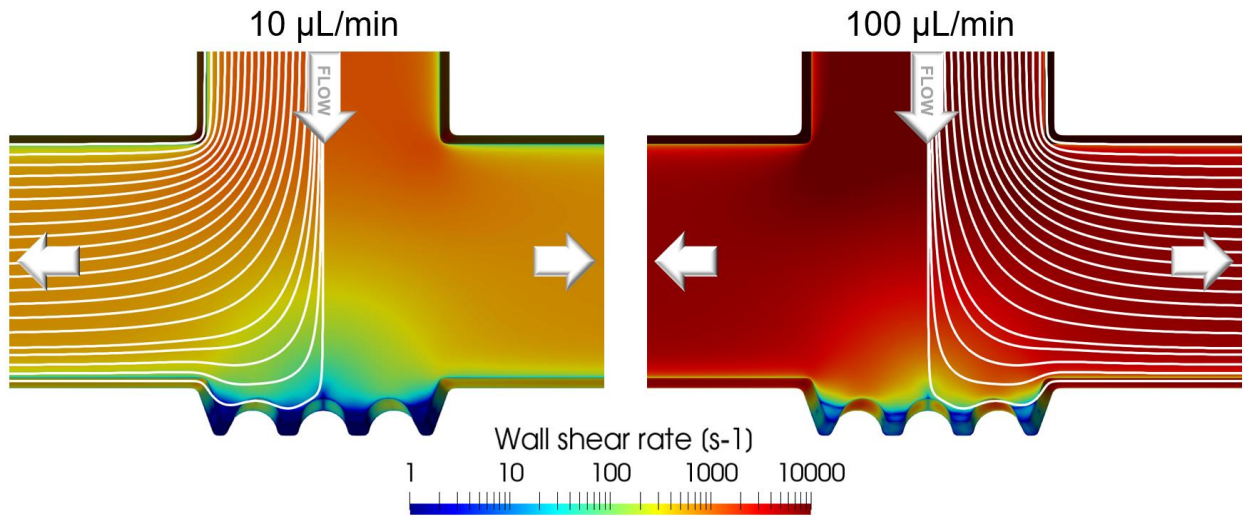

Fig. 11. Two flow rate values in the stagnation point flow channel produce an order of magnitude difference in wall shear rate. Velocity streamlines are shown on a half of the channel for visual clarity.

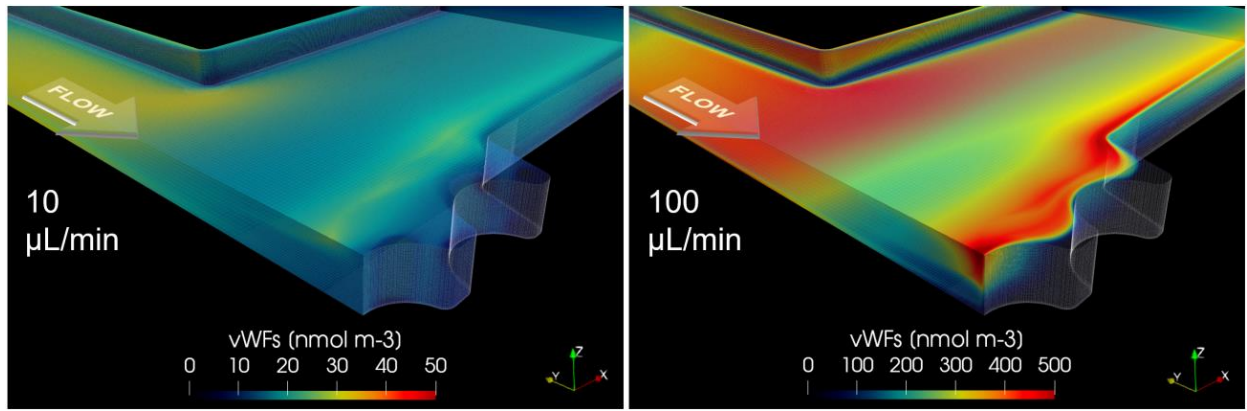

Fig. 12. Stretched vWF concentration at low and high flow rates, shown on a quarter-symmetric portion of the channel. To visualize the 3-D concentration field, the lower [vWF$_s$] visualization threshold was set to 10 nmol m$^{-3}$ and opacity mapping was enabled.

Fig. 13 presents the thrombosis simulation results compared to experiments of Herbig & Diamond (57). The insets show thrombi formed in vitro after 8 minutes of whole blood perfusion. The inset images in the left column correspond to control, and images in the right column correspond to blood treated with *N*-acetylcysteine to block the platelet-vWF interactions. In simulations, blocking the platelet-vWF interactions was achieved by setting $\alpha_{vWF} = 1$ in Eqs. (16)-(17). This disabled the [vWF$_s$]-dependent amplification of platelet deposition rates and diminution of shear cleaning rates.

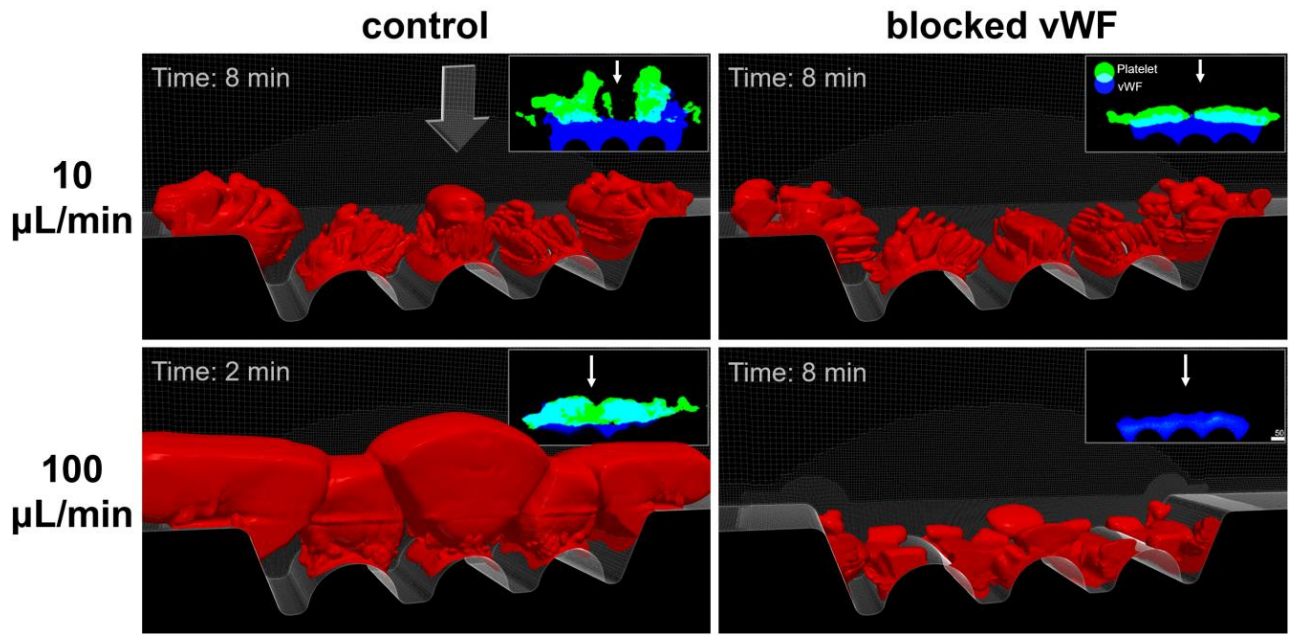

Fig. 13. Thrombus growth at 10 and 100 μL/min inlet flow rates with control and blocked platelet-vWF interactions. Simulations reproduced the distinct morphology of thrombi observed in vitro (inset): dendritic/branched clots at 10 μL/min; smooth/connected clots at 100 μL/min. Blocking the platelet-vWF interactions reduced the clot size in the lower flow rate case and drastically diminished the thrombus growth in the higher flow rate case. *Insets*: Experimental images of thrombi at 8 minutes from (57), where left column corresponds to control and right column corresponds to blood treated with *N*-acetylcysteine to block the platelet-vWF interactions. (Green signal is platelets, blue signal is vWF.) Reprinted with permission from Elsevier, *Cellular and Molecular Bioengineering*, "Thrombi Produced in Stagnation Point Flows Have a Core–Shell Structure", Herbig & Diamond, Copyright 2017 Springer Nature.

Blocking the platelet-vWF interactions in the numerical model reduced the clot size in the low flow rate case and drastically diminished the thrombus growth in the high flow rate case. In the latter, platelet deposition was eliminated from the well-washed regions and shifted into the low-shear flow recirculation zones between the cylindrical posts. (These flow features can be seen in Fig. 11.)

The morphology of thrombi formed in low and high flow rate simulations were markedly different. (See Fig. 13.) At 10 μL/min, platelets grew into dendritic structures extending normal to the flow direction. In contrast, thrombi formed at 100 μL/min appeared smoother and more connected. This result closely reproduces the experimental observations of Herbig & Diamond who described analogous morphological differences. However, our numerical model overpredicted the thrombus growth rate in the high flow rate case, hence the snapshot at 2-minute mark, while other images correspond to 8 minutes of simulation/experiment. This could be attributed to the absence of aggregate embolization in the model, whereas in vitro thrombi at high flow rates would be prone to bulk embolization after reaching a certain size.

## DISCUSSION

According to the classic Virchow's triad – originally defined for venous thrombosis and later extended to artificial surfaces – thrombosis results from concurrence of blood stasis, hypercoagulability, and a prothrombotic surface. However, arterial thrombosis is distinct from the coagulation-driven clotting in that it occurs at elevated shear rates and displays rapid kinetics driven by platelet-vWF interactions. Consequently, Casa & Ku proposed an *alternative triad* for high-shear thrombosis: (i) prothrombotic surface, (ii) platelets and vWF at sufficient concentrations, (iii) pathologically high shear rates for vWF unfolding (5). The current study presented here can be considered an extension of model of Wu et al. (33), previously validated for low to moderate shear rates (100 – 1000 $s^{-1}$) but lacking major components of the alternative triad necessary for thrombosis in high-shear conditions. This deficit was corrected in the current work by incorporating our recently presented continuum model of vWF unfolding (32).

Both the classic Virchow's triad and the alternative high-shear triad comprises a prothrombotic surface. In the numerical model of thrombosis, adhesion of platelets to a boundary surface and all the other species reactions at a boundary were modeled by surface-flux BCs, where the platelet deposition rates and the shear cleaning rates were substrate-specific coefficients. We had previously performed the calibration of these coefficients for collagen (34) and used the same values (Table 2) in all validation cases presented in this work. Importantly, the reactive BCs for platelet deposition were not applied selectively to limit the location of the thrombus growth but followed the in vitro protocol. In the stenotic tube, the reactive BCs started 4 mm upstream of the apex of the stenosis and extended downstream to the outlet of the domain. In the PFA-100®, the entire membrane surface was included. Thrombosis occurred locally *only* when the prothrombotic surface coincided with the other components of the triad: sufficient concentration of platelets and, critically, flow-induced unfolding of vWF.

In all simulations, vWF was introduced at the inlet of the domain as [$vWF_c$] – its collapsed, inert conformation. [$vWF_c$] was converted into the stretched [$vWF_s$] via either shear tumbling or strong unfolding, depending on the local extensional/rotational characteristics of the flow. [$vWF_s$] induces a local thrombogenic effect by mediating the two competing mechanisms in the model: platelet deposition and simultaneous clearing of deposited platelets by shear stress. Hence, [$vWF_s$] amplifies the deposition rates and reduces the shear cleaning rates, which tips the balance towards thrombosis thus completing the high-shear triad.

A growing thrombus is comprised of deposited platelets in resting form, [$RP_d$], and activated form, [$AP_d$]. The activation state of these platelets governs the stages of thrombus growth. In the stenotic tube, during the initial phase called the *lag time* (5), platelets accumulating near the apex of the stenosis are predominantly [$RP_d$]. However, once the combined concentration of agonists (ADP, $TxA_2$, and thrombin) reaches a threshold value, bulk activation of deposited platelets occurs, which triggers the next stage of thrombus growth called *rapid platelet accumulation* (RPA). This entire process is demonstrated in Fig. S4. When plotting the thrombus volume against time, the two phases of growth were clearly identifiable and closely matched the growth dynamics observed in vitro (Fig. 7). In the PFA-100® simulation, however, the lag time was almost non-existent (Fig. 9) because of the early platelet activation by the ADP released from the membrane surface in the collagen-ADP cartridge.

Disruption of the platelet/vWF arm of the alternative triad led to partial or complete hemostatic deficiency in the PFA-100® case. We performed simulations reflecting thrombocytopenia (1/3 platelet count), vWD Type 1 (50% vWF concentration), and vWD Type 3 (no vWF). The simulated closure times closely matched the reported PFA-100® data for these conditions (Table 5). Although the thrombocytopenia case had the longest closure time, the final thrombus volume was comparable to the baseline (Fig. 9). In contrast, the volume of the occluding clot in vWD Type 1 was almost twice that of the baseline. This difference highlights the role of vWF in this system. As the aperture shrank in diameter, the shear stress on the thrombus lumen grew rapidly, increasing the shear cleaning rate of platelets. Despite the baseline platelet count in vWD Type 1, platelets deposition within the aperture was impaired due to lack of vWF. As a result, a significant part of the growing thrombus volume was deposited outside the aperture. In the extreme case of vWD Type 3, where no vWF was present, platelets could only deposit in the low-shear zone outside the aperture (Fig. 10).

Similarly, disabling the platelet-vWF interactions in the stagnation point flow channel yielded different responses in the low and high flow rate cases (Fig. 13). In the lower flow rate case, the clot size was reduced but the location and structure of the clot were unaffected since shear rates were moderate. In the higher flow rate case, platelet deposition was eliminated from the well-washed areas and shifted into the low-shear flow recirculation zones between the cylindrical posts.

Thrombus growth is a highly complex process spanning multiple spatial and temporal scales: blood flow and transport of biochemical species, cellular-level adhesion and activation processes, and molecular level coagulation reactions. Moreover, a growing thrombus alters the local hemodynamics which feeds back into the above processes, making it a nonlinear system. Modeling this system requires balancing the complexity of the mathematical model and its breadth of application. As such, the modeling approach adopted would depend on the question being asked. For example, investigators who employ highly detailed models with numerous coagulation reactions are capable of accurately probing the specific roles of individual factors in quiescent plasma or within a defined reaction zone adjacent to injury (35, 71–75). On the other hand, a continuum CDR approach allows macroscale modeling of thrombosis in complex geometries at a reasonable computational cost (33, 41, 76–80). Particle-based methods can capture the stochastic nature of platelet deposition, heterogenous thrombus structure, and possible embolization (26, 81–85), whereas prediction of viscoelastic response of thrombus to flow is more amenable to a phase-field approach (86, 87). Recently, Shankar et al. combined a fully spatially resolved 3-D model with neural network driven platelet signaling, which enabled investigation of patient-specific platelet phenotypes (27). A comprehensive review of thrombosis modeling methods can be found in (88–91).

Our motivation in continually developing a multi-constituent model of thrombosis has been to build a numerical predictive tool for thrombus formation in full-scale medical devices. We have been incrementally improving the fidelity and versatility of the model by adding important hemodynamic and biochemical pathways that are most relevant to blood-wetted devices (33, 34, 45, 79, 80). The latest addition, presented herein, is the mechanism of vWF-mediated high-shear thrombosis. This has introduced two new species to the model, [vWF$_c$] and [vWF$_s$], and also a number of associated assumptions and parameters that carry an inherent uncertainty (92). This in turn motivated us to conduct an uncertainty quantification (UQ) study, reported by Méndez Rojano et al. (54), which elucidated areas where additional calibration and validation would yield the most improvement in accuracy and robustness of the model.

In its current state, a limitation of this model of vWF-mediated thrombosis is the omission of vWF adsorption and self-association mechanisms (93, 94). Implementing these processes would require describing the tension-dependent activation of vWF domains, association kinetics, and possible mechanisms of inhibition and self-regulation. Molecular-level simulation methods lend themselves well to problems of this kind. For example, Chen & Alexander-Katz were capable of demonstrating the reversible aggregation of polymer-colloid composites – representing vWF and platelets – using the fluctuating Lattice Boltzmann method (95). Liu et al. developed a multiscale computational model based on a coupled Lattice-Boltzmann and Langevin-dynamics method where the suspension dynamics and interactions of individual platelets and vWF multimers were resolved directly (85). Another limitation in our model is the omission of the role of platelet-released vWF. Although the effect on initial platelet adhesion should be negligible (2, 96), additional release of vWF from activated platelets could affect the thrombus growth dynamics. It must be noted that the thrombus resistance term used in Eq. (2) is similar to a Brinkman term which is valid for low volume fraction of solids and may lose detailed predictive value as the volume fraction of solids approaches 1 (97, 98). The hindered diffusion of the scalar species inside the thrombus was neglected and the diffusivity values were assumed constant. Finally, heterogeneous composition of the thrombus and its structural remodeling during growth as well as its viscoelastic response to flow were not considered.

In summary, the multi-constituent model of thrombosis presented in this work enables macro-scale 3-D simulations of thrombus formation in complex geometries over a wide range of shear rates and accounts for qualitative and quantitative hemostatic deficiencies in patient blood. Our results also demonstrate the utility of the continuum model of vWF unfolding that could be adapted to other numerical models of thrombosis.

## DECLARATION OF INTEREST
The authors declare no competing interests.

## ACKNOWLEDGMENTS
Author Wei-Tao Wu thanks the support of the Natural Science Foundation of Jiangsu No. BK20201302. This work was supported by the National Institutes of Health grant R01HL089456.

# SUPPORTING MATERIAL

# von Willebrand Factor unfolding mediates platelet deposition in a model of high-shear thrombosis

Mansur Zhussupbekov[1], Rodrigo Méndez Rojano[1], Wei-Tao Wu[2], James F. Antaki[1,*]

[1] Meinig School of Biomedical Engineering, Cornell University, Ithaca, NY, USA
[2] Department of Aerospace Science and Technology, Nanjing University of Science and Technology, Nanjing, China
*Correspondence: antaki@cornell.edu

## SUPPLEMENTARY METHODS
### Governing equations of chemical/biological species
Provided below are the full set of governing equations in expanded form.

**Unactivated resting PLTs in flow ([RP])**

$$\frac{\partial [RP]}{\partial t} + div(v_f[RP]) = div(D_P \nabla [RP]) - k_{apa}[RP] - k_{spa}[RP] - k_{rpd}[RP] + f_{emb,r}[RP_d] \quad (18)$$

where $v_f$ is the velocity of the fluid (blood), $D_P$ is the diffusivity of the platelets, $k_{apa}$ and $k_{spa}$ are the platelets activation rates due to agonists and shear stress, respectively, and $k_{rpd}$ is the deposition rate between [RP] and deposited platelets. (For detailed mathematical definition and numerical implementation of the $k_{rpd}$ term see the next section.) $f_{emb,r}$ is the cleaning rate [RP$_d$] due to shear stress.

**Activated platelets in flow [AP]**

$$\frac{\partial [AP]}{\partial t} + div(v_f[AP]) = div(D_P \nabla [AP]) + k_{apa}[RP] + k_{spa}[RP] - k_{apd}[AP] + f_{emb,a}[AP_d] \quad (19)$$

where $k_{apd}$ is the deposition rate between [AP] and deposited platelets. (For detailed mathematical definition and numerical implementation of the $k_{apd}$ term see the next section.) $f_{emb,a}$ is the cleaning rate [AP$_d$] due to shear stress.

**Deposited resting platelets [RP$_d$]**

$$\frac{\partial [RP_d]}{\partial t} = (1-\theta)k_{rpd}[RP] - k_{apa}[RP_d] - k_{spa}[RP_d] - f_{emb,r}[RP_d] \quad (20)$$

where $\theta$ is the fraction of [RP] activated by contact with biomaterial surfaces and deposited platelets.

**Deposited activated platelets [AP$_d$]**

$$\frac{\partial [AP_d]}{\partial t} = \theta k_{rpd}[RP] + k_{apd}[AP] + k_{apa}[RP_d] + k_{spa}[RP_d] - f_{emb,a}[AP_d] \quad (21)$$

**PLT-released agonist (ADP) [a$_{pr}$]**

$$\begin{aligned}\frac{\partial [a_{pr}]}{\partial t} + div(v_f[a_{pr}]) &= div(D_{apr}\nabla[a_{pr}]) \\ &\quad + \lambda_j \big(k_{apa}[RP] + k_{spa}[RP] + k_{apa}[RP_d] + k_{spa}[RP_d] + \theta k_{rpd}[RP]\big) \\ &\quad - k_{1,j}[a_{pr}]\end{aligned} \quad (22)$$

where $\lambda_j$ is the amount of agonist $j$ released per platelet and $k_{1,j}$ is the inhibition rate constant of agonist $j$; in this case, $j$ represents ADP. Therefore, $\lambda_j k_{apa}[RP]$ and $\lambda_j k_{apa}[RP_d]$ are the ADP release attributed to the activation of resting platelets due to agonist, $\lambda_j k_{spa}[RP]$ and $\lambda_j k_{spa}[RP_d]$ are the ADP release attributed to the activation of resting platelets due to shear stress, and $\lambda_j \theta k_{rpd}[RP]$ represents the ADP release attributed to the platelets activation due to contact. $k_{1,j}[a_{pr}]$ is the inhibition rate of ADP.

**PLT-synthesized agonist (TxA₂) [$a_{ps}$]**

$$\frac{\partial [a_{ps}]}{\partial t} + div(v_f[a_{ps}]) = div(D_{aps}\nabla[a_{ps}]) + s_{pj}([AP] + [AP_d]) - k_{1,j}[a_{ps}] \tag{23}$$

where $s_{pj}$ is the rate constant of synthesis of the agonist $j$; in this case, $j$ represents TxA₂. Therefore, $s_{pj}[AP]$ and $s_{pj}[AP_d]$ represent the rate of synthesis of TxA₂ due to activated platelets in flow and deposited activated platelets. $k_{1,j}[a_{ps}]$ is the inhibition rate of TxA₂.

**Prothrombin [PT]**

$$\frac{\partial [PT]}{\partial t} + div(v_f[PT]) = div(D_{PT}\nabla[PT]) - \varepsilon[PT]\left(\phi_{at}([AP] + [AP_d]) + \phi_{rt}([RP] + [RP_d])\right) \tag{24}$$

where $\varepsilon$ is the unit conversion from NIH units to SI units, $\phi_{at}$ and $\phi_{rt}$ are the thrombin generation rate constant on the surface of activated platelets and unactivated resting platelets. Therefore $\phi_{at}[AP]$ and $\phi_{rt}[RP]$ are the thrombin generation rate due to activated and resting platelets in flow, whereas $\phi_{at}[AP_d]$ and $\phi_{rt}[RP_d]$ are the thrombin generation rate from deposited activated and resting platelets, respectively.

**Thrombin [TB]**

$$\frac{\partial [TB]}{\partial t} + div(v_f[TB]) = div(D_{TB}\nabla[TB]) + [PT]\left(\phi_{at}([AP] + [AP_d]) + \phi_{rt}([RP] + [RP_d])\right) - \Gamma[TB] \tag{25}$$

where $\Gamma$ is the Griffith's template model for the kinetics of the heparin-catalyzed inactivation of thrombin by ATIII. Therefore $\Gamma[TB]$ is the inactivation rate of thrombin by ATIII.

**ATIII [AT]**

$$\frac{\partial [AT]}{\partial t} + div(v_f[AT]) = div(D_{AT}\nabla[AT]) - \Gamma\varepsilon[TB] \tag{26}$$

where $\Gamma\varepsilon[TB]$ is the consumption rate of ATIII due to inactivation of thrombin.

## Mathematical and Numerical Considerations for Deposition Rates *k*<sub>rpd</sub> and *k*<sub>apd</sub>.

Provided below are mathematical and numerical considerations regarding the two terms in equations (1) – (4) above, that represent the rates of deposition of resting (unactivated) platelets, $k_{rpd}$, and activated platelets, $k_{apd}$, to the surface of the thrombus. Referring to the finite volume schematic depicted in Figure S1, the $k_{rpd}$ of the finite-volume-cell 5 is calculated as

$$k_{rpd} = div(k_{pd,f}\vec{n})k_{ra} \tag{27}$$

where $k_{ra}$ is a constant, $\vec{n}$ is the normal to the finite-volume-face, and $f$ refers to the faces shared by adjacent cells 2, 4, 6 and 8. Assuming that thrombus grows layer by layer, when the volume fraction of the deposited platelets of a finite-volume cell $\phi$ is greater than a critical value $\phi_c$ (for example, 0.74, which is the maximum packing fraction for spheres), this cell is able to influence the neighbor cells. Therefore,

$$k_{pd,f} = \begin{cases} \dfrac{[AP_d]_f}{PLT_{max}}, & \phi > \phi_c \\ 0, & \phi < \phi_c \end{cases} \quad (28)$$

where $\dfrac{[AP_d]_f}{PLT_{max}}$ represents the percentage of the area occupied by deposited activated platelets at that mesh face. This rule does not apply to the boundary faces.

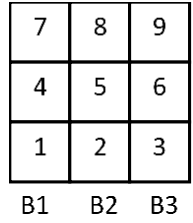

**Fig S1.** Schematic of cells and faces of a mesh.

## Variables, coefficients, and parameters of the thrombosis model

**Table S1.** Source terms associated with platelets.

| Species | $[C_i]$ abbreviation | $S_i$ form |
|---|---|---|
| Unactivated Resting PLTs | [RP] | $-k_{apa}[RP] - k_{spa}[RP] - k_{rpd}[RP] + f_{emb,r}[RP_d]$ |
| Activated PLTs | [AP] | $k_{apa}[RP] + k_{spa}[RP] - k_{apd}[AP] + f_{emb,a}[AP_d]$ |
| Deposited Resting PLTs | [RP$_d$] | $(1-\theta)k_{rpd}[RP] - k_{apa}[RP_d] - k_{spa}[RP_d] - f_{emb,r}[RP_d]$ |
| Deposited Activated PLTs | [AP$_d$] | $\theta k_{rpd}[RP] + k_{apd}[AP] + k_{apa}[RP_d] + k_{spa}[RP_d] - f_{emb,a}[AP_d]$ |

**Table S2.** Source terms for chemical species of the model.

| Species | $[C_i]$ abbreviation | $S_i$ form |
|---|---|---|
| PLT-released agonist (ADP) | [a$_{pr}$] | $\lambda_j\big(k_{apa}[RP] + k_{spa}[RP] + k_{apa}[RP_d] + k_{spa}[RP_d] + \theta k_{rpd}[RP]\big) - k_{1,j}[a_{pr}]$ |
| PLT-synthesized agonist (TxA$_2$) | [a$_{ps}$] | $s_{pj}([AP] + [AP_d]) - k_{1,j}[a_{ps}]$ |
| Prothrombin | [PT] | $-\varepsilon[PT]\big(\phi_{at}([AP] + [AP_d]) + \phi_{rt}([RP] + [RP_d])\big)$ |
| Thrombin | [TB] | $-\Gamma[TB] + [PT]\big(\phi_{at}([AP] + [AP_d]) + \phi_{rt}([RP] + [RP_d])\big)$ |
| ATIII | [AT] | $-\Gamma\varepsilon[TB]$ |

**Table S3.** Species units, coefficient of species diffusion and initial condition. $\gamma$ is the local shear rate. For more detail see Sorenson(1, 2) and Goodman (3).

| Species | Species units | $D_i (m^2 s^{-1})$ | Initial (inlet) condition in blood |
|---|---|---|---|
| [RP] | PLT m$^{-3}$ | $1.58 \times 10^{-13} + 6.0 \times 10^{-13} \gamma$ | $1.5 \times 10^{14} - 4.5 \times 10^{14}$ (Human) |
| [AP] | PLT m$^{-3}$ | $1.58 \times 10^{-13} + 6.0 \times 10^{-13} \gamma$ | $0.01[RP] - 0.05[RP]$ |
| [a$_{pr}$] | nmol m$^{-3}$ | $2.57 \times 10^{-10}$ | 0.0 |
| [a$_{ps}$] | nmol m$^{-3}$ | $2.14 \times 10^{-10}$ | 0.0 |
| [PT] | nmol m$^{-3}$ | $3.32 \times 10^{-11}$ | $1.1 \times 10^6$ |
| [TB] | Um$^{-3}$ | $4.16 \times 10^{-11}$ | 0.0 |
| [AT] | nmol m$^{-3}$ | $3.49 \times 10^{-11}$ | $2.844 \times 10^6$ |
| [RP$_d$] | PLT m$^{-3}$ | N/A | 0.0 |
| [AP$_d$] | PLT m$^{-3}$ | N/A | 0.0 |

**Table S4.** Species boundary conditions.

| Species [$C_i$] | $j_i$ form | Description |
|---|---|---|
| [RP] | $-Sk_{rpd,b}[RP] + f_{embb,r}[RP_d]$ | Consumption due to [RP]-surface adhesion; Generation due to shear cleaning of [RP$_d$]. |
| [AP] | $-Sk_{apdb}[AP] + f_{embb,a}[AP_d]$ | Consumption due to [AP]-surface adhesion; Generation due to shear cleaning of [AP$_d$]. |
| [a$_{pr}$] | $\lambda_j(k_{apa}[RP_d] + k_{spa}[RP_d] + \theta Sk_{rpdb}[RP])$ | Generation due to agonist and shear activation of [RP$_d$]; Generation due to surface contact activation of [RP]-surface adhesion. |
| [a$_{ps}$] | $s_{pj}[AP_d]$ | Platelet-synthesized generation due to [AP$_d$] |
| [PT] | $-\varepsilon[PT](\phi_{at}[AP_d] + \phi_{rt}[RP_d])$ | Consumption due to thrombin, [TB], generation. |
| [TB] | $[PT](\phi_{at}[AP_d] + \phi_{rt}[RP_d])$ | Generation from prothrombin, [PT], due to deposited platelets. |
| [AT] | 0.0 | No reaction flux. |

**Table S5.** Species boundary conditions. (See Table S6 definitions of each of the terms.)

| Species [$C_i$] | $j_i$ form | Description |
|---|---|---|
| [$RP_d$] | $\int_0^t \left((1-\theta)Sk_{rpd,b}[RP] - k_{apa}[RP_d] - k_{spa}[RP_d] - f_{embb,r}[RP_d]\right) dt$ | Generation due to [RP]-surface adhesion; Consumption due to agonist and shear activation; Consumption due to shear cleaning. |
| [$AP_d$] | $\int_0^t \left(Sk_{apd,b}[AP] + \theta Sk_{rpdb}[RP] + k_{apa}[RP_d] + k_{spa}[RP_d] - f_{embb,a}[AP_d]\right) dt$ | Generation due to [AP]-Surface adhesion; Generation due to surface contact activation of [RP]-Surface adhesion; Generation due to agonist and shear activation of [RP]; Consumption due to shear cleaning. |

**Table S6.** Value/expression and description of reaction terms and parameters.

| Terms | Value or expression | units | Description |
|---|---|---|---|
| $\phi$ | $\dfrac{RP_d + AP_d}{PLT_{max}}$ | (N/A) | Volume fraction of deposited platelets (thrombus). |
| $k_{apa}$ | $\begin{cases} 0, \Omega < 1.0 \\ \dfrac{\Omega}{t_{ct}}, \Omega \geq 1.0 \\ \dfrac{1}{t_{act}}, \dfrac{\Omega}{t_{ct}} \geq \dfrac{1}{t_{act}} \end{cases}$ | ($s^{-1}$) | Platelets activation due to agonists; $t_{ct}$ is the characteristic time, which can be used for adjusting the activation rate, and here we choose $t_{ct} = 1s$ as provided Sorensen (1, 2); $\left(k_{apa} = \dfrac{1}{t_{act}}, \dfrac{\Omega}{t_{ct}} \geq \dfrac{1}{t_{act}}\right)$ implies the reaction cannot be faster than platelets physical activation procedure and 99% platelets will be activated during the activation procedure if the agonists or shear stress is large enough. $t_{act}$ is the characteristic time; $t_{act}$ is suggested to range from 0.1s to 0.5s considering the results by Frojmovic et. al (4) and Richardson (5). |
| $\Omega$ | $\sum_{j=1}^{n_a} w_j \dfrac{a_j}{a_{j,crit}}$ | (N/A) | $a_j$ refers to the concentration of ADP, TxA$_2$ and Thrombin. For the value of $w_j$ and $a_{j,crit}$, see Table S7. |
| $k_{spa}$ | $\dfrac{1}{t_{ct,spa}} = \begin{cases} \dfrac{1}{4.0 \times 10^6 \tau^{-2.3}}, t_{ct,spa} > t_{act} \\ \dfrac{1}{t_{ct,spa}}, t_{ct,spa} < t_{act} \end{cases}$ | ($s^{-1}$) | Platelet activation due to shear stress, $\tau$. Expression $t_{ct,spa} = 4.0 \times 10^6 \tau^{-2.3}$ was adopted from Goodman(3) and Hellums(6). |
| $k_{rpd}$ | $k_{rpd} = \begin{cases} div(k_{pd,f}\vec{n})k_{ra}, \phi < 1 \\ 0, \phi = 1 \end{cases}$ | ($s^{-1}$) | Deposition rate between [RP] and deposited platelets. $f$ refers to the face of a mesh cell and $\vec{n}$ is the unit normal to the face. |

| Symbol | Expression/Value | Units | Description |
|---|---|---|---|
| | | | Deposition is turned off if the thrombus volume fraction value, $\phi$, reaches 1. |
| | | | For mathematical details on how this term is calculated see the previous section. |
| $k_{ra}$ | $3.0 \times 10^{-6}$ | (m s$^{-1}$) | Constant related to $k_{rpd}$ (1, 2). |
| $k_{apd}$ | $k_{apd} = \begin{cases} div(k_{pd,f}\vec{n})k_{aa}, \phi < 1 \\ 0, \phi = 1 \end{cases}$ | (s$^{-1}$) | Deposition rate between [AP] and deposited platelets. Deposition is turned off if the thrombus volume fraction value, $\phi$, reaches 1. For mathematical details on how this term is calculated see the previous section. |
| $k_{aa}$ | $3.0 \times 10^{-5}$ | (m s$^{-1}$) | Constant related to $k_{apd}$ (1, 2). |
| $f_{emb,r}$ | $Dia_{PLT} \, div(k_{emb,f}\vec{n})\left(1 - exp(-0.0095\frac{\tau}{\tau_{rd}})\right)$ | s$^{-1}$ | [RP$_d$] shear cleaning (embolization due to shear stress). Expression $exp(-0.0095\tau)$ was suggested by Goodman (3). $Dia_{PLT} = 2.78 \times 10^{-6} m$ is the hydraulic diameter of platelets. |
| $f_{emb,a}$ | $Dia_{PLT} \, div(k_{emb,f}\vec{n})\left(1 - exp(-0.0095\frac{\tau}{\tau_{ad}})\right)$ | s$^{-1}$ | [AP$_d$] shear cleaning (embolization due to shear stress) (3). |
| $k_{emb,f}$ | $k_{emb,f} = \begin{cases} 1, k_{pd,f} > 0 \\ 0, k_{pd,f} = 0 \end{cases}$ | s$^{-1}$ | Activation of shear cleaning in a cell during thrombus propagation. See previous section for details. |
| $\tau_{rd}$ | 15 | $dyne \, cm^{-2}$ | [RP$_d$] shear cleaning related constant in $f_{emb,r}$. |
| $\tau_{ad}$ | 30 | $dyne \, cm^{-2}$ | [AP$_d$] shear cleaning related constant in $f_{emb,a}$. |
| $f_{embb,r}$ | $\left(1 - exp(-0.0095\frac{\tau}{\tau_{rd,b}})\right)$ | s$^{-1}$ | [RP$_d$] shear cleaning (embolization due to shear stress) in boundary due to shear stress. |
| $f_{embb,a}$ | $\left(1 - exp(-0.0095\frac{\tau}{\tau_{ad,b}})\right)$ | s$^{-1}$ | [AP$_d$] shear cleaning (embolization due to shear stress) in boundary due to shear stress. |
| $\tau_{rd,b}$ | To be determined, depends on bio-material. | $dyne \, cm^{-2}$ | [RP$_d$] shear cleaning related constant in boundary in $f_{embb,r}$. |
| $\tau_{ad,b}$ | To be determined, depends on bio-material. | $dyne \, cm^{-2}$ | [AP$_d$] shear cleaning related constant in boundary in $f_{embb,a}$. |
| $PLT_{max}$ | $\frac{PLT_{s,max}}{Dia_{PLT}}$ | PLT m$^{-3}$ | The maximum concentration of platelets in space. $PLT_{s,max} = 7 \times 10^{10} PLT m^{-2}$ is the total capacity of the surface for platelets; $Dia_{PLT} = 2.78 \times 10^{-6} m$ is the hydraulic diameter of platelets. |
| $\lambda_j$ | $2.4 \times 10^{-8}$ | nmol PLT$^{-3}$ | The amount of agonist $j$ released per platelet. |
| $\theta$ | 0.1 | (N/A) | Platelet contact activation. |
| $k_{1,j}$ | $\begin{cases} 0.0161 \text{ for TxA}_2 \\ 0.0 \text{ for ADP} \end{cases}$ | $(s^{-1})$ | The inhibition rate constant of agonist $j$. |
| $s_{pj}$ | $\begin{cases} 9.5 \times 10^{-12} \text{ for TxA}_2 \\ 0 \text{ for ADP} \end{cases}$ | nmol $PLT^{-3}$ | The rate constant of synthesis of an agonist. |
| $\varepsilon$ | $9.11 \times 10^{-3}$ | nmol $U^{-1}$ | Unit conversion from NIH units to SI units. |
| $\phi_{at}$ | $3.69 \times 10^{-15}$ | $m^3$ $nmol^{-1}$ $PLT^{-1}$ $U$ $s^{-1}$ | Thrombin generation rate on the surface of activated platelets. |
| $\phi_{rt}$ | $6.5 \times 10^{-16}$ | $m^3$ $nmol^{-1}$ $PLT^{-1}$ $U$ $s^{-1}$ | Thrombin generation rate on the surface of resting (unactivated) platelets. |
| $\Gamma$ | $\dfrac{k_{1,T}[H][AT]}{\alpha K_{AT} K_T + \alpha K_{AT}\varepsilon[TB] + [AT]\varepsilon[TB]}$ | $(s^{-1})$ | Griffith's template model for the kinetics of the heparin-catalyzed inactivation of thrombin by ATIII. |
| $k_{1,T}$ | 13.333 | $(s^{-1})$ | A first-order rate constant. |
| $[H]$ | $0.1 \times 10^6$ | nmol $m^{-3}$ | Heparin concentration, assuming specific activity of 300 U $mg^{-1}$ and molecular weight of 16 kDa.(2, 7) |
| $\alpha$ | 1.0 | (N/A) | A factor to simulate a change in affinity of heparin for ATIII when it is bound to thrombin or for thrombin when it is bound to ATIII. |
| $K_{AT}$ | $0.1 \times 10^6$ | nmol $m^{-3}$ | The dissociation constant for heparin/ATIII. |
| $K_T$ | $3.5 \times 10^4$ | nmol $m^{-3}$ | The dissociation constant for heparin/thrombin. |
| $S$ | $1 - \dfrac{[RP_d]_b + [AP_d]_b}{PLT_{s,max}}$ | (N/A) | Percentage of the wall (boundary) not occupied by deposited platelets. The subscript $b$ indicates the boundary field values. |
| $k_{rpd,b}$ | To be determined, depends on bio-material. | $(m\ s^{-1})$ | Unactivated platelet-boundary(wall) deposition rate. |
| $k_{apd,b}$ | To be determined, depends on bio-material. | $(m\ s^{-1})$ | Activated platelet-boundary(wall) deposition rate. |

**Table S7.** Threshold concentration of agonists for platelet activation and agonist-specific weight(1–3). It should be noted that the agonist thresholds may vary according to different experimental studies. For ADP see (4, 8–11); for Thrombin see (12–14); for TxA$_2$ see (15–17).

| Species | $a_{j,crit}$ | $w_j$ |
|---|---|---|
| ADP ($[a_{pr}]$) | $1.00 \times 10^6$ nmol $m^{-3}$ | 1 |
| TxA$_2$ ($[a_{ps}]$) | $0.20 \times 10^6$ nmol $m^{-3}$ | 3.3 |
| Thrombin ($[TB]$) | $0.10 \times 10^6$ U $m^{-3}$ | 30 |

## Thrombus resistance term coefficient

In the thrombus resistance term in Eq. 2 of the manuscript, we have previously used $C_2 = 2 \times 10^9$ kg/(m³s) that was computed assuming the deposited platelets behave like densely compact particles 2.78um in diameter (18–20). Recently, Du et al. (21) quantified the permeability of red clots formed under stasis and white clots formed under high shear conditions. Using Darcy's law, they obtained permeability values of $5 \times 10^{-4} \pm 5 \times 10^{-4}$ μm² for red clots and $0.3 \pm 0.4$ ⁴ μm² for white clots. To compare the permeability of the clot in our numerical model of thrombosis, we conducted a simulation where a fluid with a constant viscosity, $\mu$, was forced through a thrombus of a defined length, $L$, and cross-sectional area, $A$, at a constant flow rate, $Q$ (Fig. S1). The resulting pressure drop across the clot, $\Delta P$, was used to calculate the clot permeability, $\kappa$, from the Darcy's law:

$$\kappa = \frac{\mu Q L}{A \Delta P}$$

Using the previously reported value of $C_2 = 2 \times 10^9$ kg/(m³s) in the thrombus resistance term yielded the permeability value of $\kappa = 2.35 \times 10^{-4}$ μm², which fits within the permeability range for red clots reported by Du et al. Since they found that platelet-rich white clots formed at high shear rates were three orders of magnitude more permeable than the red clots, we use $C_2 = 2 \times 10^6$ kg/(m³s) for the current study of vWF-mediated high-shear thrombosis. This value yields the permeability of $\kappa = 0.235$ μm² – similar to the white clot permeability from Du et al.

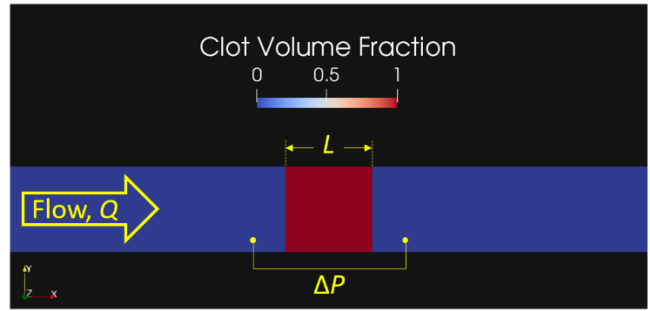

**Fig S2.** Simulation setup for calculating the clot permeability using Darcy's law.

## vWF unfolding thresholds and monomer diffusion time

Sing and Alexander-Katz report the vWF unfolding thresholds from their simulations in a nondimensionalized form using $Wi$, where for extensional flow, $\dot{\epsilon}\tau_{mon} \approx 0.316$, and for simple shear, $\dot{\gamma}\tau_{mon} \approx 5.62$ (22). They use the monomer diffusion time, $\tau_{mon}$, which is calculated as:

$$\tau_{mon} = \frac{a^2}{\mu_0 k_B T}, \qquad \mu_0 = \frac{1}{6\pi \eta a}$$

where $a$ is vWF monomer radius, $\eta$ is the solvent dynamic viscosity, $k_B$ is Boltzmann constant, and $T$ is absolute temperature in K. Therefore, the value of $\tau_{mon}$ will depend on the assumption of the monomer radius, $a$. Sing and Alexander-Katz assumed $a = 50$ nm in their prior work (23). The vWF repeating unit has been reported to be around 70 nm along its major axis (24, 25). In the interest of consistency, we used the simple shear unfolding threshold from Eq. (10), $\dot{\gamma}_{1/2} = 5522$ s⁻¹, to compute $\tau_{mon}$ from $\dot{\gamma}\tau_{mon} \approx 5.62$. To verify the resulting monomer diffusion time of $\tau_{mon} = 1.02$ ms, we compute the corresponding monomer radius $a$. Assuming blood plasma at 37°C, $\eta = 0.0012$ kg·m⁻¹s⁻¹, T = 310.15 K, the monomer radius is $a = 58$ nm, which falls within the range of values mentioned above.

## PFA-100® membrane boundary conditions for ADP

In the PFA-100® system, blood is forced through a central orifice of a bioactive membrane until occlusion is achieved. The membrane of the collagen/ADP cartridge is coated with 50 μg ADP to promote activation of platelets. Since the thrombosis model uses a volume concentration of ADP, we account for the ADP coating by prescribing a diffusive flux of $[a_{pr}]$ in the boundary condition for the membrane walls.

50 μg of ADP converts to 118 nmol using molecular weight of 427 g/mol. Dividing this by the volume of the membrane yields a concentration of $[a_{pr}] = 5.29 \times 10^{11}$ nmol m$^{-3}$.

From Fick's First Law, flux is $J = -D \frac{dC}{dx}$, where $D$ is the diffusion coefficient and $\frac{dC}{dx}$ is the concentration gradient. Using $D = 2.57 \times 10^{-10}$ m$^2$ s$^{-1}$ for ADP and $x = 1 \times 10^{-6}$ m as the distance from the boundary to the center of the first cell, the flux $J \approx 1 \times 10^8$ nmol m$^2$ s$^{-1}$. From a parametric study, we determined that near-surface platelets are already activated at $J = 1 \times 10^6$ nmol m$^2$ s$^{-1}$ and further increasing the flux has no significant effect on the occlusion time.

Since it is not known if the ADP from the membrane completely depletes during the PFA-100® test, two additional simulations were conducted to assess the effect of the constant ADP flux from the membrane. First, completely disabling the ADP flux resulted in a prolonged lag time for the initial platelet activation and thrombus growth, as can be seen from Figure S3. However, the effect on the subsequent thrombus growth rate was insignificant. To verify this, another simulation was conducted, where the ADP flux from the membrane, $J$, was active only for the first 20 seconds (baseline lag time) of the simulation:

$$J = \begin{cases} 1 \times 10^8 \text{ nmol m}^2 \text{ s}^{-1}, & t = [0, 20 \text{ s}] \\ 0, & t = [20, 300 \text{ s}] \end{cases}$$

In this case, the effect on the thrombus growth was negligible. This shows that after the initial platelet activation within the first 20 seconds, the bulk thrombus growth does not rely on the ADP from the membrane.

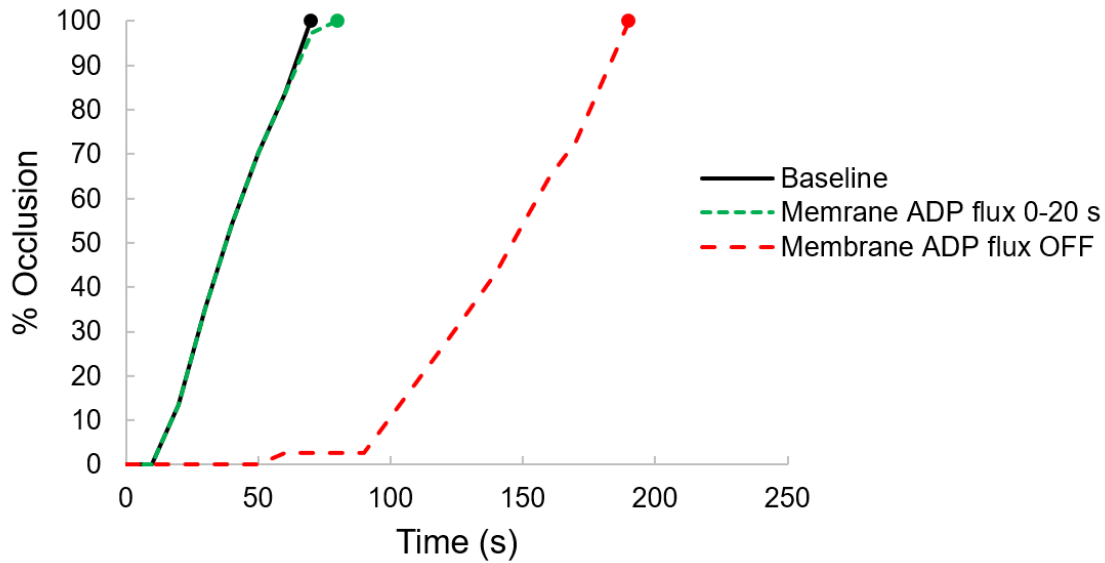

Fig. S3. The effect of ADP flux on the thrombus growth rate.

## Statistical analysis of the reported PFA-100® closure time data

Harrison et al. reported PFA-100 closure times for patients classified as normal or diagnosed with vWD or thrombocytopenia (26). Graph data points for the Collagen-ADP (CADP) cartridge from Harrison et al. were digitized using (27) and converted into numerical data. The mean and standard deviation of the closure times were calculated using NumPy module in Python.

Harrison et al. reported all vWD cases (Type 1, Type 2B, and Type 3) lumped together, but also stated that all severe vWD samples (Type 2B and Type 3) yielded closure times ≥ 300 s. That is the maximum test duration after which the test is terminated without closure of the aperture. Therefore, to calculate the mean and standard deviation of the vWD Type 1 samples, we omitted the data points with closure times ≥ 300 s. Accordingly, for vWD Type 3, we reported the closure time as ≥ 300 s.

## SUPPLEMENTARY FIGURES
### Stenotic tube

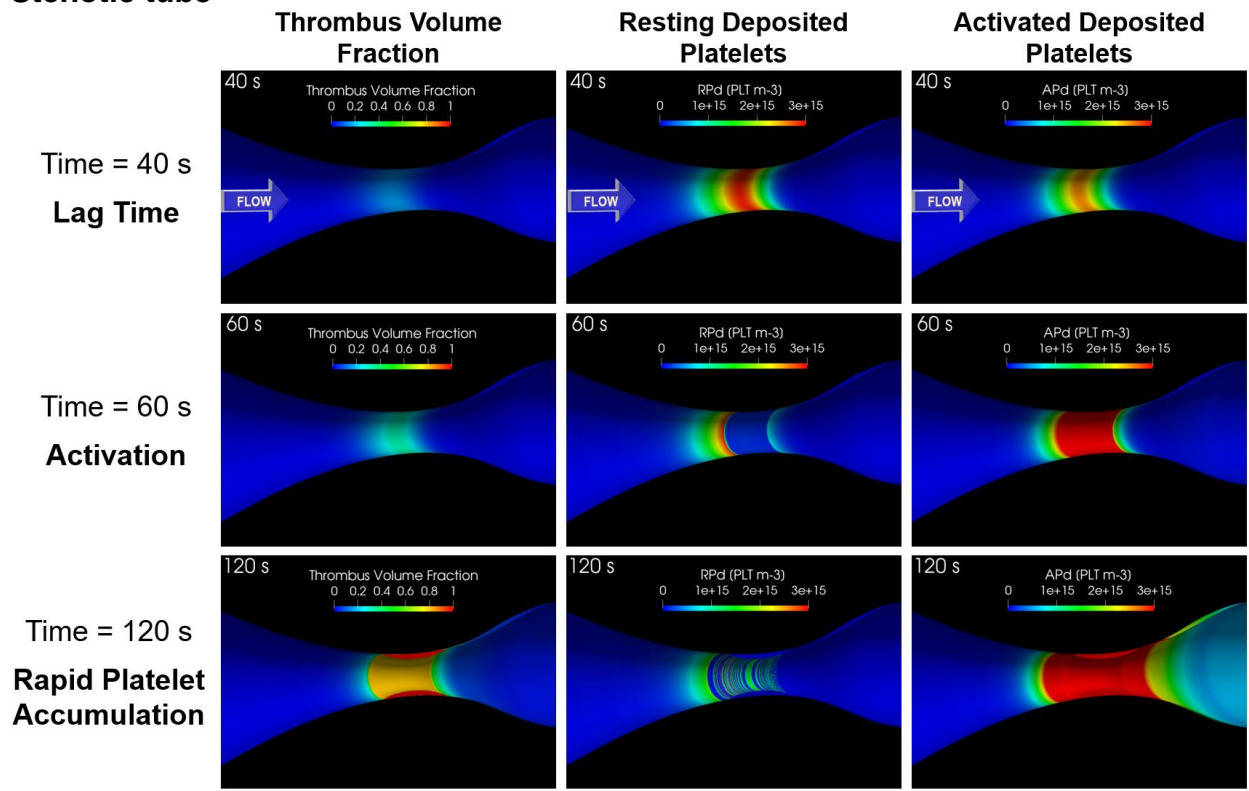

**Fig S4.** Thrombus forming at the apex of the stenotic tube is comprised of Resting (unactivated) Deposited Platelets, [$RP_d$], and Activated Deposited Platelets, [$AP_d$]. Large-scale activation of deposited platelets initiates the transition from the *Lag time* phase into the *Rapid Platelet Accumulation* (RPA) phase of thrombosis, as described by Casa & Ku (28).

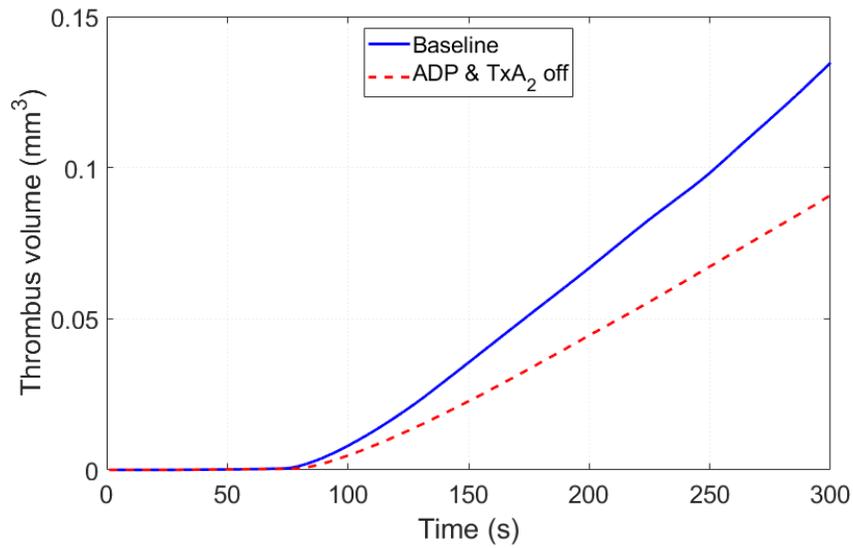

**Fig S5.** The effect of disabling the action ADP and TxA$_2$ on the thrombus growth in the stenotic tube, mimicking a dual anti-platelet therapy.

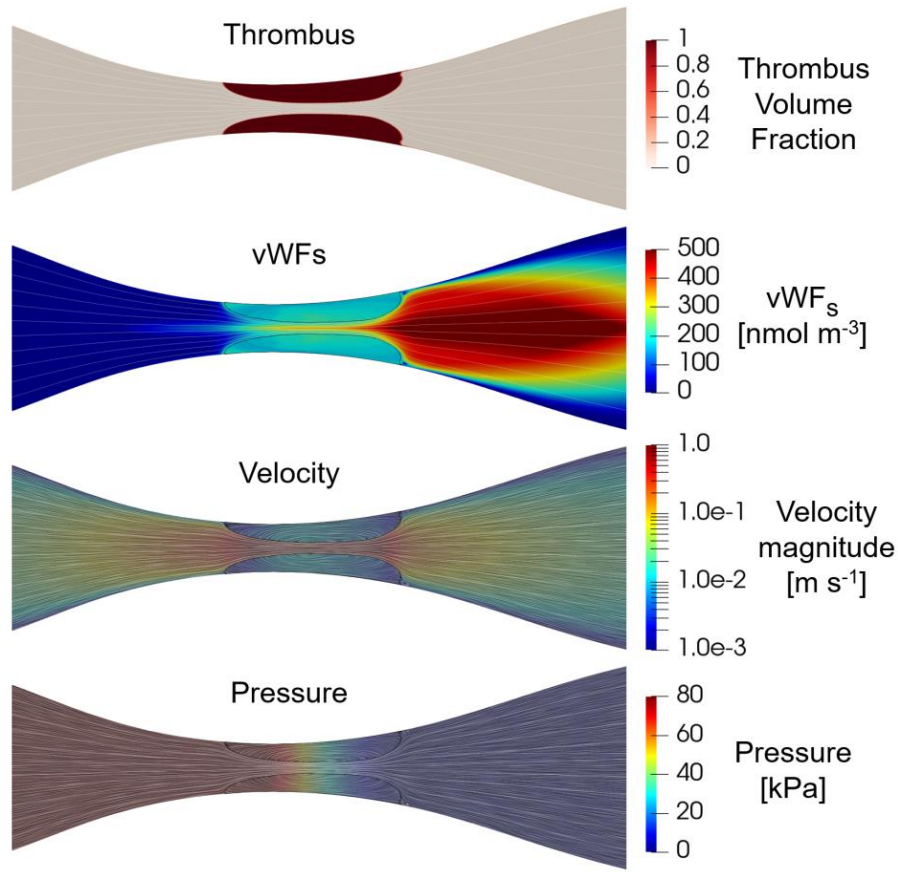

**Fig S6.** The flow and vWF$_s$ concentration inside the thrombus formed at the apex of the stenotic tube at t = 240 s. Note the log scale for the velocity magnitude. The large pressure gradient across the stenosis forces the flow in and out of the thrombus. The Peclet number for vWF is $Pe \gg 1$, indicating that vWF transport within the clot is dominated by the convective transport.

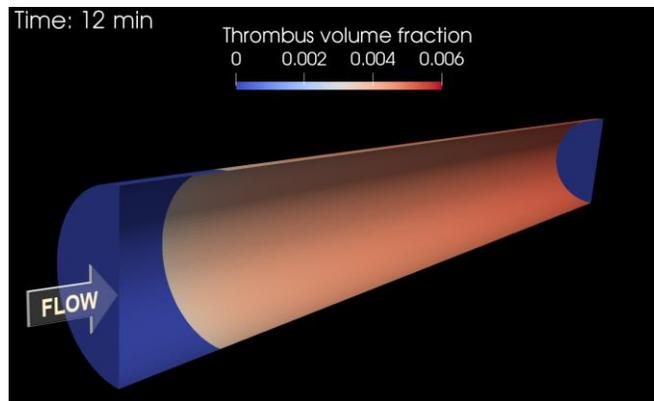

**Fig S7.** Simulation of blood perfusion through a straight collagen-coated tube at a wall shear rate of 1000 s$^{-1}$. After 12 minutes of simulation, platelets accumulated only in the first layer of mesh cells. The simulation reproduced the results of a control experiment conducted by Bark, Para, and Ku, where they observed only a thin monolayer of platelets after 12 minutes of blood perfusion at a wall shear rate of 1000 s$^{-1}$ (29).

## SUPPLEMENTARY REFERENCES